\documentclass[twocolumn,tighten]{aastex7}

\usepackage{graphicx}
\usepackage{xcolor}
\definecolor{dblue}{rgb}{0.0, 0.0, 0.55}
\usepackage[version=4]{mhchem}
\usepackage[normalem]{ulem}

\def\Vl{V_{\!\ell}}

\def\nHim1{n_{\langle\rm H\rangle,i-1}}
\def\nHip1{n_{\langle\rm H\rangle,i+1}}
\def\vdreq{{v^{\hspace{-0.8ex}^{\circ}}_{\rm dr}}}

\def\Sr{S_{r}}
\def\vth{v_{\rm th}}

\def\vdreq{{v^{\hspace{-0.8ex}^{\circ}}_{\rm dr}}}

\def\St{{S\hspace*{-0.1ex}t}}

\def\Coll{\Delta{\rm Coll}}

\def\aap{A\&A}

\def\apjl{ApJL}
\def\apjs{ApJS}

\def\la{\mathrel{\mathchoice {\vcenter{\offinterlineskip\halign{\hfil
$\displaystyle##$\hfil\cr<\cr\sim\cr}}}
{\vcenter{\offinterlineskip\halign{\hfil$\textstyle##$\hfil\cr
<\cr\sim\cr}}}
{\vcenter{\offinterlineskip\halign{\hfil$\scriptstyle##$\hfil\cr
<\cr\sim\cr}}}
{\vcenter{\offinterlineskip\halign{\hfil$\scriptscriptstyle##$\hfil\cr
<\cr\sim\cr}}}}}
\def\ga{\mathrel{\mathchoice {\vcenter{\offinterlineskip\halign{\hfil
$\displaystyle##$\hfil\cr>\cr\sim\cr}}}
{\vcenter{\offinterlineskip\halign{\hfil$\textstyle##$\hfil\cr
>\cr\sim\cr}}}
{\vcenter{\offinterlineskip\halign{\hfil$\scriptstyle##$\hfil\cr
>\cr\sim\cr}}}
{\vcenter{\offinterlineskip\halign{\hfil$\scriptscriptstyle##$\hfil\cr
>\cr\sim\cr}}}}}

\newcommand{\ms}[1]{\textcolor{blue}{#1}}

\begin{document} 

\title{\Large{Prediction of sulphate hazes in the lower Venus atmosphere}}

\author{Peter~Woitke}
  \affiliation{Space Research Institute, Austrian Academy of Sciences,
               Schmiedlstr.~6, A-8042, Graz, Austria}
  \email{peter.woitke@oeaw.ac.at}
\author{Manuel~Scherf}
  \affiliation{Space Research Institute, Austrian Academy of Sciences,
               Schmiedlstr.~6, A-8042, Graz, Austria}
  \email{manuel.scherf@oeaw.ac.at}
\author{Christiane~Helling}
  \affiliation{Space Research Institute, Austrian Academy of Sciences,
               Schmiedlstr.~6, A-8042, Graz, Austria}
  \affiliation{Institute for Theoretical Physics and Computational
               Physics, Graz University of Technology, Petersgasse 16, 
               8010 Graz, Austria}
  \email{christiane.helling@oeaw.ac.at}
\author{Paul~Rimmer}
  \affiliation{Department of Earth Sciences, University of Cambridge,
               Downing Street, Cambridge CB2 3EQ, UK}
  \affiliation{Cavendish Laboratory, University of Cambridge, JJ Thomson
               Avenue, Cambridge CB3 0HE, UK}
  \affiliation{MRC Laboratory of Molecular Biology, Francis Crick Avenue,
               Cambridge CB2 0QH, UK}
  \email{pbr27@cam.ac.uk}
\author{Martin~Ferus}
  \affiliation{J. Heyrovsk{\'y} Institute of Physical Chemistry, Czech
               Academy of Sciences, Prague, Czech Republic}
  \email{martinferus@email.cz}
\author{Helmut~Lammer}
  \affiliation{Space Research Institute, Austrian Academy of Sciences,
               Schmiedlstr.~6, A-8042, Graz, Austria}
  \email{helmut.lammer@oeaw.ac.at}
\author{Fabian~Weichbold}
  \affiliation{Space Research Institute, Austrian Academy of Sciences,
               Schmiedlstr.~6, A-8042, Graz, Austria}
  \email{fabian.weichbold@oeaw.ac.at}
\author{Kate\v{r}ina~N\v{e}me\v{c}kov{\'a}}
  \affiliation{J. Heyrovsk{\'y} Institute of Physical Chemistry, Czech
               Academy of Sciences, Prague, Czech Republic}
  \email{nemeckova.kata@gmail.com}
\author{Petr~Eminger}
  \affiliation{J. Heyrovsk{\'y} Institute of Physical Chemistry, Czech
               Academy of Sciences, Prague, Czech Republic}
  \affiliation{Department of Physical and Macromolecular Chemistry,
               Faculty of Science, Charles University, Prague,
               Czech Republic}
  \email{petr.eminger@jh-inst.cas.cz}
\author{Jaroslav Kačina}
  \affiliation{J. Heyrovsk{\'y} Institute of Physical Chemistry, Czech
               Academy of Sciences, Prague, Czech Republic}
  \affiliation{Institute of Geochemistry, Mineralogy and Mineral
               Resources, Faculty of Science, Charles University,
               Prague, Czech Republic}
  \email{jaroslav.kacina@jh-inst.cas.cz}
\author{Tereza Constantinou}
  \affiliation{Institute of Astronomy, University of Cambridge,
               Madingley Road, Cambridge CB3 0HA, UK}
  \email{tc496@cam.ac.uk}
  
\begin{abstract}\noindent
  We study the amount, size distribution and material composition of
  (sub-)$\mu$m aerosol particles in the lower Venus atmosphere
  $<50\,$km. Our GGchem phase-equilibrium model predicts
  metal-chloride and metal-fluoride molecules to be present in the gas
  over the Venus surface in trace concentrations $<\!2\times10^{-12}$,
  in particular FeCl$_2$, NaCl, KCl and SiF$_4$.  Using an improved
  version of the DiffuDrift model developed by \cite{Woitke2020}, we
  find that these molecules deposit to form solid potassium sulphate
  K$_2$SO$_4$, sodium sulphate Na$_2$SO$_4$, and pyrite FeS$_2$ above
  about 15.5\,km, 9.5\,km and 2.4\,km, respectively. These heights
  coincide well with the three potential haze layers found in the
  Pioneer Venus Large Probe neutral mass spectrometer data by
  \cite{Mogul2023}.  The particles with radius $<0.3\,\mu$m can be
  dredged up from the ground to reach the sulphuric acid cloud base
  from below by diffusion. The particle density decreases from
  $\sim 5000\rm\,cm^{-3}$ at ground level to $\sim 100\rm\,cm^{-3}$ at a
  height of 45\,km. Particles larger than about 1\,$\mu$m are found to
  stay confined to the ground $<\!10\,$km, indicating that the larger,
  so-called mode~3 particles, if they exist, cannot originate from the
  surface. All particles are expected to be coated by a thin layer of
  FeS$_2$, Na$_2$SO$_4$ and K$_2$SO$_4$.  We have included the
  repelling effect of particle charges on the coagulation, without
  which the model would predict much too steep gradients close to the
  surface, which is inconsistent with the measured opacities.  Our
  models suggest that the particles must have at least {100 negative charges per micron of particle radius at ground level, and $>50/\mu$m} at a height of 45\,km.\\*[-10mm]
\end{abstract}


%-------------------------------------------------------------------
\section{Introduction}
%-------------------------------------------------------------------
\noindent
The exploration of Earth's inner neighbour planet, Venus, with the
Venera {\citep[e.g.,][]{Moroz1983,Moroz2002}, Mariner \citep[e.g.,][]{Sonett1963,Kliore1967,Fjeldbo1971},
Pioneer Venus \citep[e.g.,][]{Knollenberg1980,Oyama1980,Colin1980}, Vega
\citep[e.g.,][]{Sagdeev1986,Sagdeev1986b,Bertaux1996}, Venus Express \citep{Svedhem2007,Titov2009}, Akatsuki
(Venus Climate Orbiter) \citep{Nakamura2016}, and Magellan
\citep[e.g.,][]{Saunders1992}} space missions included flybys, descent probes,
orbiters, and balloons. These missions revealed the basic physical and
chemical structure of the Venus atmosphere, clouds and aerosols, its
surface, and geology {\citep[see, e.g.,][for further reviews on Venus]{Hunten1983,Bougher1997,Fegley2014}}.

The atmosphere of Venus is a chemically exotic and complex system, {see
e.g.\,\cite{deBergh2006}, \cite{Titov2018} and \cite{Dai2025}}, involving photochemistry in the upper
atmosphere, thermochemical cycles in the troposphere, and presumably a
strong buffering role of the surface {for various molecules like CO$_2$, HCl, HF and some S-bearing species \citep[e.g.,][]{Fegley1992,Zolotov2018,Zolotov2024}}, which tends
to establish thermochemical equilibrium in the gas in direct contact
with the surface \citep{Rimmer2021}. 
{Earlier works about the buffering mechanisms include e.g.\,\citet{Krasnopolsky1994, Krasnopolsky2007}, and \cite{Fegley1997}. A recent summary of various equilibrium models for the surface of Venus can be found in Table 16.2 of \cite{Zolotov2024}.} 

Venus has a thick atmosphere that contains gaseous and particle 
absorbers responsible for a powerful greenhouse effect that results 
in a mean surface temperatures of about 735\,K, {with substantial variations with day/night, latitude, and elevation, see e.g. \cite{Bertaux2007}}. 
Contrary to Earth and Mars, Venus absorbs a significant
fraction of the solar energy in the optically thick clouds that
surround the planet {at heights of about (43-47)\,km to (63-75)\,km, depending on latitude \citep{Zolotov2024}}. The cloud
tops include an unknown ultraviolet (UV) absorber
\citep[e.g.,][]{Zasova1981,Titov2018}, for which the dimer of ferric
chloride \ce{FeCl3} \citep{Krasnopolsky1985,Krasnopolsky2017}, two
isomers of \ce{S2O2} \citep{Frandsen2016}, rhomboclase and ferric sulphates \citep{Jiang2024}, or even
microbial lifeforms \citep[e.g.,][]{Grinspoon1997,Limaye2018} were
suggested as potential absorbing agents.

%Thermochemical cycles involving water vapour, sulphur and carbon
%species most likely dominate in this region.
In the upper clouds, sulphuric acid is identified as a particulate
component, although the chemical origin and physical processes in the
cloud deck are still not fully understood {\citep[see, e.g., reviews by][]{Esposito1983,Esposito1997,Marcq2018,Titov2018}}. 
The strong depletion of SO$_2$ within the clouds
\citep[e.g.,][]{Vandaele2017a,Vandaele2017b} remains a
conundrum, although hydroxide salts in the cloud droplets were
recently shown to provide a potential sink for SO$_2$
\citep{Rimmer2021}.

The study of aerosol particles in the Venus atmosphere provides
critical insight into the planetary environment and has profound
implications for the understanding of atmospheric chemistry, cloud
formation, and the possibility of the existence of an aerial life
cycle. The Pioneer Venus {Orbiter cloud photopolarimeter \citep{Kawabata1980} and descent probes \citep{Knollenberg1980}} measured the density and size distribution of aerosol particles in and below the clouds, revealing a possibly tri-modal size
distribution with mean radii of $\sim$\,0.2\,$\mu$m, (mode 1),
$\sim$\,1.0\,$\mu$m (mode 2) and $\sim$\,3.6\,$\mu$m (mode 3). The
latter have been suggested to have a crystalline structure
\citep{Knollenberg1984,Carlson1993}, possibly coated with sulphuric
acid \citep{Grinspoon1993}, which would indicate growth within the
clouds. However, the existence of mode~3 particles has been challenged
and suggested to be the tail-end of mode 2 rather than an independent
particle population \citep{Toon1984}. The latter also fits the data
from the {Soviet} descent probes which did not observe a separate
mode~3 particle distribution \citep{Moshkin1986,Zasova1996}. The
non-existence of the mode~3 particles is further supported by
\citet{Gao2014}, who used the community aerosol and radiation model
for atmospheres (CARMA) \citep{Turco1983} to interpret the PICAV/SOIR
data from Venus Express. In their model, they found a bimodal
distribution, where the mode~3 particles are merely a larger, grown
version of the mode~2 particles that evaporate below the clouds.

The particle size distribution in the Venusian clouds is compatible
with the known range of cells and spores of microorganisms on
Earth. Based on this, \citet{Seager2021} suggested an aerial life
cycle around the base of the Venus liquid sulphuric acid clouds, where
desiccated spores (comparable to mode 1 particles) populate the layers
below the clouds, diffuse up into the clouds where they are
incorporated in the cloud droplets and germinate to a metabolically
active life form. The droplets further grow by coagulation until they
reach droplet sizes comparable to mode 2 particles and start to settle
gravitationally, evaporate and release new
spores. \citet{Bains2021Mode3} suggested another aerial life cycle
where metabolically active microbes in cloud droplets produce NH$_3$
which raises the droplet $p$H, traps SO$_2$ and H$_2$O in sulphate
salts and form large semisolid mode 3 particles. These particles again
settle gravitationally below the cloud deck, evaporate and release
spores, SO$_2$ and H$_2$O, a process that could also explain the
abovementioned SO$_2$ depletion in the Venus clouds. Another study by
\citet{SchulzeMakuch2004} links the mode 3 particles in the lower
atmosphere of Venus with life by proposing that mode 3 particles are
microbes coated in elemental sulphur that float through the dense gas
of the atmosphere \citep[see also,][]{Seager2021}. The astrobiological
potential of the Venus atmosphere, the chemical anomalies, and the
unexplained cloud properties have been recently reviewed by
\cite{Petkowski2024}.

Phosphine, potentially synthesised by microorganisms, was tentatively
detected in the clouds of Venus at parts-per-billion-levels
\citep{Greaves2021}.  By now, no abiotic sources are known that could
sufficiently explain the potential existence of PH$_3$ in the upper
Venus atmosphere \citep{Bains2024Phosphine}.  The influx of
micrometeorites \citep{Gao2014} is too little to, even if all
phosphorous was converted into phosphine \citep{Bains2021Phosphine},
and volcanism \citep{Truong2021} could only produce the
ppb-level signals under extreme volcanic activity
\citep{Bains2022Phosphine}.  Abiotic photochemical pathways for
phosphine synthesis on acidic mineral surfaces have been suggested by
\cite{Mrazikova2024}, but are not yet experimentally confirmed.
However, similar photochemical pathways on mineral surfaces have been
experimentally verified for the synthesis of \ce{CH4}
(methanogenese) from \ce{CO2} by \cite{Civis2016} and
\cite{Civis2019}. Such processes could mimic biosignature-like redox
disequilibria through abiotic chemistry.

Understanding the amount, size distribution and material composition
of aerosol particles is a crucial step toward evaluating their ability
to influence the redox balance in the gas through hydride synthesis
and thereby contribute to chemical anomalies. Therefore, aerosols are
not merely a passive component of the atmosphere but can actively
shape Venus' chemical dynamics.  Another pivotal question is whether
these particles can reach the clouds from below to seed the clouds,
but by now only a few studies investigated the chemical and diffusive
behaviour of such aerosol particles in the lower Venusian atmosphere.

Simulations and measurements of aerosol particles in the lowest parts
of Venus’ atmosphere are scarce. \citet{Anderson1969} developed a
simple model to estimate the density and maximum particle size of dust
at altitudes between 0 and 25\,km. Based on Mariner~5 and Venera~4
data, they found that for the same convective activity as on Earth,
more dust will be in Venus’ atmosphere than on Earth with diameters as
large as 130\,$\mu$m. \citet{Sagan1975} found that particles of sizes
up to $40\,\mu$m can be lifted from the surface when the threshold
frictional velocity is $>1-2\rm\,cm/s$, corresponding to a horizontal
wind velocity of $>\!0.3\,$m/s above the surface boundary layer. If
these particles reach a height of 10\,km, they could be redistributed
around the entire planet, as Venera~8 measured wind velocities of a
few $10\,$m/s at these heights. {Experiments in a Venus Wind Tunnel by \citet{Greeley1984} further suggest that dust grains below 40\,$\mu$m can be lifted by the wind and float in the Venusian atmosphere \citep{Greeley1984,Greeley1990}, which is in good agreement with the calculations performed by \citet{Sagan1975}. See also \citet{Carter2023} for a review on aeolian transport.}

Continuum absorption at 1.18\,$\mu$m seen in the deep atmosphere of
Venus by the VIRTIS and SPICAV-IR instruments onboard Venus Express
could indeed relate to the presence of dust particles
\citep{Snels2014}. {The existence of layers of dust and/or haze in the lower atmosphere of Venus are consistent with a variety of observations. Early measurements by Mariner~V and Venera~4 were interpreted as dust layer at $\la\!25$\,km \citep{Anderson1969}. The detection of small backscatter signals in the Pioneer Venus night and north probes around an altitude of $\sim\!6$\,km was suggested to indicate a particle-bearing layer \citep{Seiff1995}.}
Later analysis of the Pioneer Venus and Venera
descent probes further support the existence of low-altitude haze
layers.  A reconstruction of the CO$_2$ profile measured by the 
Pioneer Venus Large Probe neutral mass spectrometer indicates the
presence of haze layers below an altitude of 17\,km, in particular 
at $\sim\!15\pm2$\,km, $\sim\!10\pm3$\,km, and $\sim\!3\pm1$\,km,
related to partial clogging of the instrument by particles
\citep{Mogul2023}, an effect that was even stronger in the main
sulphuric acid clouds. A reanalysis of spectrophotometer \citep{Grieger2004} and discharge
current measurements \citep{Lorenz2018} on Venera 13 and 14 also
support a dusty atmosphere below 35\,km.  Based on the discharge
current data measured at 1-2\,km altitudes, \cite{Lorenz2018} found a
near-surface charge density of the order of $\rm 1000\,pC/m^3$.
Assuming about singly charged particles, and a particle size of
$\sim\!0.3\,\mu$m, \citet{Lorenz2018} suggested a density of charged
particles of order $1000\rm\,cm^{-3}$, which would correspond to an
aerosol opacity of order $\rm 1/km$, backing the earlier results of
\cite{Grieger2004}.  These charge and particle densities are higher
than observed for the Saharan dust at high altitudes on Earth
\citep{Nicoll2011}. 
However, these large aerosol opacities seem to conflict with the surface pictures taken by the Venera landers {\citep{Selivanov1983,Pieters1986}\footnote{See also \url{http://mentallandscape.com/C_CatalogVenus.htm} for the Venera lander images digitally remastered by Don Mitchel.}}, which showed a clear horizon. This could indicate that there are indeed much fewer particles that are charged much more, such that the resulting charge density remains the same while the opacity is lower.

{A recent study by \citet{Kulkarni2025}, which reconstructed spectrophotometric data from the Venera~13 mission by digitizing old graphic material, supports the existence of a near-surface particulate layer. Its peak density was found to be at an altitude of 3.5-5\,km \citep[in agreement with the secondary peak found by][]{Grieger2004} with a log-normal particle size distribution peaking at particle radii between 0.6 and 0.85\,$\mu$m. The refractory index of the particles was derived to be $\eta_{\rm\,r} \sim 1.4 - 1.6$, which closely matches $\eta_{\rm\,r}\sim 1.52$ of basalt particles. This indicates that the particle layer could have formed from basaltic dust lifted from the surface \citep{Kulkarni2025}.}

A lower haze layer is also consistent with thermo-chemical equilibrium
simulations by \citet{Byrne2024} using the GGchem model
\citep{Woitke2018} for the Venus setup as published in
\cite{Rimmer2021}, which indicates the formation and rainout of pyrite
clouds above an altitude of about 3\,km, an effect that could explain
the high reflectivity of the Venus mountain tops seen in radar
measurements \citep{Pettengill1982,Pettengill1996}, {because pyrite is a conducting material. Pyrite was considered by \cite{Klose1992} and \cite{Kohler2015}, together with other potential materials, to explain the radar reflectivity, see also \cite{Barsukov1982, Barsukov1986,Zolotov1992}}. Alternatively, the formation of heavy metal frost {through the condensation of volcanically degassed volatile metals, such as Cu, As, Pb, Sb, and Bi, at higher elevations was first proposed by \citet{Brackett1995} and later refined by \citet{Schaefer2004}}. {Volume scattering from a potentially porous, highly weathered soil layer was suggested as an additional alternative \citep[e.g.,][]{Pettengill1992,Tryka1992,Brackett1995}}. For a review of the oxidation state of the lower atmosphere and surface of Venus, see \cite{Fegley1997}; \ms{for a review of its surface mineralogy, see \citet{Gilmore2023}}.

%\textcolor{red}{\bf from Martin}: 
%Venus' atmosphere serves as a unique natural laboratory
%for studying photochemical processes, particularly those occurring on
%aerosols. These processes can mimic biosignature-like redox
%disequilibria through fully abiotic chemistry, as summarized in our
%previous study \cite{Mrazikova2024}. The photochemical reduction of
%oxides on aerosol surfaces has been experimentally demonstrated, such
%as in the formation of methane (\ce{CH4}) from carbon dioxide
%(\ce{CO2}) on mineral particle surfaces. This mechanism has also been
%theoretically explored in the context of the recent detection of
%phosphine (\ce{PH3}) in Venus' clouds.

Understanding the chemical inventory of Venus' aerosols is essential
for determining how specific chemical compounds influence the
thermochemistry of the planet's atmosphere and interact with other
atmospheric species. The current study represents a crucial step
toward estimating the chemical nature of the aerosol particles within Venus' atmospheric profile.

The paper is organised as follows. In Sect.\,\ref{sec:GGchem} we
describe the results obtained with our thermo-chemical equilibrium
code {\sc GGchem}.  We discuss (i) the concentrations found for
metal-chloride and metal-fluoride molecules over the Venus surface,
and (ii) the stability of condensates in the Venus atmosphere up to
a height of 50\,km.  In Sect.\,\ref{sec:DiffuDrift}, we
introduce the {\sc DiffuDrift~v2} code to model chemically active aerosol
particles including gravitational settling, diffusion and coagulation.
We assume that a mixture of the condensates found to be stable in
Sect.\,\ref{sec:GGchem} can deposit on these particles, and use a
kinetic description of the process of deposition and sublimation to
predict the chemical composition, number density and  sizes of
these particles.  Section~\ref{sec:results} then discusses the results
of four simulations, (i) a basic model of passive aerosol particles,
(ii) a model for passive particles including coagulation, (iii) a
model for chemically active particles based on the second model,
and (iv) a model for chemically active particles with a reduced
abundance of the passive particles at $z=0$. Section~\ref{sec:summary}
summarises our results.

\begin{table*}
  \caption{Trace concentrations of metal-chloride and metal-fluoride
    molecules over the Venus surface and first cloud condensates
    predicted by our {\sc GGchem} model.}
  \label{tab:conc}
  \hspace*{-28mm}\resizebox{208mm}{!}{\begin{tabular}{rcl|c|c|c|rcl}
    \hline
    && main & trace & first cloud & cloud &&& \\[-1mm]
    \!\!element \hspace*{-4mm} && \hspace*{-4mm} molecule
    & \!concentration$^{(1)}\!\!$
    & condensate  & \!base [km]\!\!
    & hypothetical && \hspace*{-8mm} surface reaction \\
    \hline
    &&&&&&\\[-3ex]           
    Fe &$\to$& \ce{FeCl2} & $1.3\times10^{-12}$ & \ce{FeS2[s]}$^{(2)}$ & 2.9 &
      \ce{FeCl2} + 2 \ce{SO2} + \ce{H2O} + 5 \ce{CO}
      &\!\!\!$\to$& \ce{FeS2[s]} + 2  \ce{HCl} + 5  \ce{CO2}\\
    Na &$\to$& \ce{NaCl } & $1.1\times10^{-12}$ & \ce{Na2SO4[s]} & 9.3 &
      2 \ce{NaCl} + \ce{SO2} + \ce{H2O} + \ce{CO2}
      &\!\!\!$\to$& \ce{Na2SO4[s]} + 2 HCl + CO\\
    Si &$\to$& \ce{SiF4} & $7.0\times10^{-13}$ & -- & -- \\
    K  &$\to$& \ce{KCl} & $5.2\times10^{-13}$   & \ce{K2SO4[s]} & 13.8 &
      2 \ce{KCl} + \ce{SO2} + \ce{H2O} + \ce{CO2}
      &\!\!\!$\to$& \ce{K2SO4[s]} + 2 HCl + CO\\
    Al &$\to$& \ce{AlF2O} & $6.4\times10^{-14}$ & \ce{Al2O3[s]}  & 0.0 &
      2 \ce{AlF2O} + 2 \ce{H2O} + \ce{CO}
      &\!\!\!$\to$& \ce{Al2O3[s]} + 4 \ce{HF} + \ce{CO2} \\
    Ti &$\to$& \ce{TiF4 } & $1.4\times10^{-18}$ & \ce{TiO2[s]}   & 0.0 &
      \ce{TiF4} + 2 \ce{H2O}
      &\!\!\!$\to$& \ce{TiO2[s]} + 4 \ce{HF} \\
    Mg &$\to$& \ce{MgCl2} & $8.5\times10^{-21}$ & \ce{MgF2[s]}   & 0.0 &
      \ce{MgCl2} + 2 \ce{HF}
      &\!\!\!$\to$& \ce{MgF2[s]} + 2 \ce{HCl} \\
    Ca &$\to$& \ce{CaCl2} & $1.4\times10^{-21}$ & \ce{CaSO4[s]}  & 0.0 &
      \ce{CaCl2} + \ce{SO2} + \ce{H2O} + \ce{CO2}
      &\!\!\!$\to$& \ce{CaSO4[s]} + 2 \ce{HCl} + \ce{CO} \\
   % &&&&&&\\[-2.2ex]           
   % \hline
   % &&&&&&\\[-2.2ex]           
   %  S  &$\to$& \ce{SO2  } & $ 150\times10^{-6}$ \\
   %  H  &$\to$& \ce{H2O  } & $  30\times10^{-6}$ \\
   %  Cl &$\to$& \ce{HCl  } & $0.50\times10^{-6}$ \\
   %  F  &$\to$& \ce{HF   } & $0.35\times10^{-6}$ \\
    \hline
  \end{tabular}}\\[1mm]
  \footnotesize
  $^{(1)}$: molecular particle concentrations with respect to
  the total gas particle density $n_j/n$\\
  $^{(2)}$: the notation [s] means solid, and [l] liquid. {Species without square brackets are gas species.}\\*[-2mm]
\end{table*}

%-------------------------------------------------------------------
\section{The GGchem model}
%-------------------------------------------------------------------
\label{sec:GGchem}
\noindent
{We first summarise our previously obtained results for the composition of the gas and the material composition of the surface when applying the phase-equilibrium model {\sc GGchem} \citep{Woitke2018} to the foot-point of the Venus atmosphere, see \citet{Rimmer2021} for more details. {\sc GGchem} predicts these properties based on the surface pressure, surface temperature, and total (condensed $+$ gas phase) element abundances. In order to determine these total element abundances, we considered} 
a combination of (i) the surface oxide ratios measured by the Vega 2 lander
\citep{Surkov1986} and (ii) a set of measured molecular concentrations 
in the lower Venus atmosphere. The included molecular concentrations 
have been compiled from various missions and instruments as summarised 
in \citep{Rimmer2021}: \ce{CO2}, \ce{N2}, \ce{SO2}, \ce{H2O}, \ce{OCS}, 
\ce{CO}, \ce{HF}, \ce{HCl}, \ce{H2S}, \ce{S3}, \ce{S4}, and \ce{NO}.  After carefully adjusting the total oxygen abundance, the {\sc GGchem}
equilibrium condensation model is capable of approximately reproducing
both the element composition of the solid Venus surface and the gas
phase composition over its surface. All molecules that are predicted to
be abundant in our model (those with percent or ppm concentrations)
have observed counterparts and their concentrations agree within a
factor of 2 or better with the measurements. Other molecules, such as
\ce{O2}, \ce{CH4} and \ce{NH3}, have concentrations $<\!10^{-15}$ in
our model and are indeed not observed, {see table 3 in
\citet{Rimmer2021}.}

{Therefore, we concluded in \cite{Rimmer2021} that the near-surface atmosphere of Venus is in fact close to gas phase chemical and phase equilibrium with its hot surface, in agreement with \citet{Zolotov1996}, \citet{Krasnopolsky2013}, and \citet{Bierson2020}, and that our simple {\sc GGchem} equilibrium condensation model can be confidently used to interpret the measurements.}

{The second part of table 3 in \citet{Rimmer2021} lists the 11 condensates found to be stable along with their mass fractions found in our model for the Venus surface:
\ce{MgSiO3[s]} {\sl (enstatite)},
\ce{CaAl2Si2O8[s]} {\sl (anorthite)},
\ce{NaAlSi3O8[s]} {\sl (albite)},
\ce{Fe2O3[s]} {\sl (hematite)},
\ce{CaSO4[s]} {\sl (anhydrite)},
\ce{SiO2[s]} {\sl (quartz)},
\ce{Al2O3[s]} {\sl (corundum)},
\ce{KAlSi3O8[s]} {\sl (microcline)},
\ce{Mn3Al2Si3O12[s]} {\sl (spessartine)},
\ce{TiO2[s]} {\sl (rutile)}, and
\ce{MgF2[s]} {\sl (magnesium ﬂuoride)}. These condensates are all saturated ($S=1$) in the model, and present in the surface, whereas all other condensates are undersaturated $S<1$ and not present in the surface.  This is a consequence of our simple phase equilibrium model that includes only pure condensates. More sophisticated models include solid solutions \citep[e.g.][]{Barsukov1982,Barsukov1986} and discuss the presence of \ce{Fe3O4} {\sl (magnetite)} and \ce{Al2SiO5[s]} {\sl (andalusite)}, see e.g. \cite{Fegley1992} and \cite{Zolotov2024}. Indeed, these two minerals obtain supersaturation ratios of 0.95 and 0.93 in our model, respectively, so they are on the verge of becoming stable in our model.}

{Microscopically speaking, the applicability of chemical and phase equilibrium means that there must be a local balance between sublimation and re-sublimation rates, and there must be sufficiently rapid chemical processes in the gas phase to install that equilibrium over the surface.} 

{Table~\ref{tab:conc} shows new details from our previously published phase equilibrium model for the Venus surface. The first two columns show the main molecular carriers of Fe, Na, Si, K, Al, Ti, Mg and Ca in the gas phase, and their particle concentrations over the surface. We mainly find those element carriers to be chlorine and fluorine molecules, which is in agreement with e.g. \cite{Schaefer2004b}.  We note that \cite{Zolotov2021, Zolotov2025} argue that chloride molecules would decompose on the surface when in contact with solid sulphates, however in our model, the surface is too hot to host \ce{Na2SO4[s]} and \ce{K2SO4[s]} as stable condensates.}

Noteworthy, the main Fe-bearing molecule in our model is iron dichloride (\ce{FeCl2}) but not the dimer of ferric chloride (\ce{Fe2Cl6}) as suggested by \citep{Krasnopolsky2017}.  Our model does include {\ce{FeCl3} and \ce{Fe2Cl6}, with thermo-chemical data from the NIST/Janaf tables \citep{Chase1998}, but the concentrations of these molecules result to be 4 and 15 orders of magnitudes lower than the concentration of \ce{FeCl2}}, respectively, which itself is more than $10^4\times$ lower than the estimate of 17\,ppb of \ce{Fe2Cl6} by Krasnopolsky.
{According to our model, if iron-chloride molecules were indeed present in ppb-concentrations, they should immediately deposit on the surface of the aerosol particles.}
{Additional iron molecules of potential interest are \ce{FeF3}, \ce{Fe(OH)2}, \ce{Fe2Cl4} and \ce{FeF2}, which are found to have concentrations between $10^{-14}$ and $10^{-21}$.}

%             nFECL2 1.143E+09 1.288E-12
%              nFEF3 6.636E+06 7.483E-15
%             nFECL3 3.350E+05 3.778E-16
%           nFE(OH)2 9.155E+03 1.032E-17
%            nFE2CL4 7.840E+02 8.840E-19
%              nFEF2 2.616E+00 2.950E-21
%          n(FECL3)2 8.993E-06 1.014E-26

\begin{figure*}
  \vspace*{-2mm}
  \centering
  \begin{tabular}{ccc}
  \hspace*{-2mm}
  \includegraphics[page=2,height=100mm,trim=0 0 41 0,clip]
                  {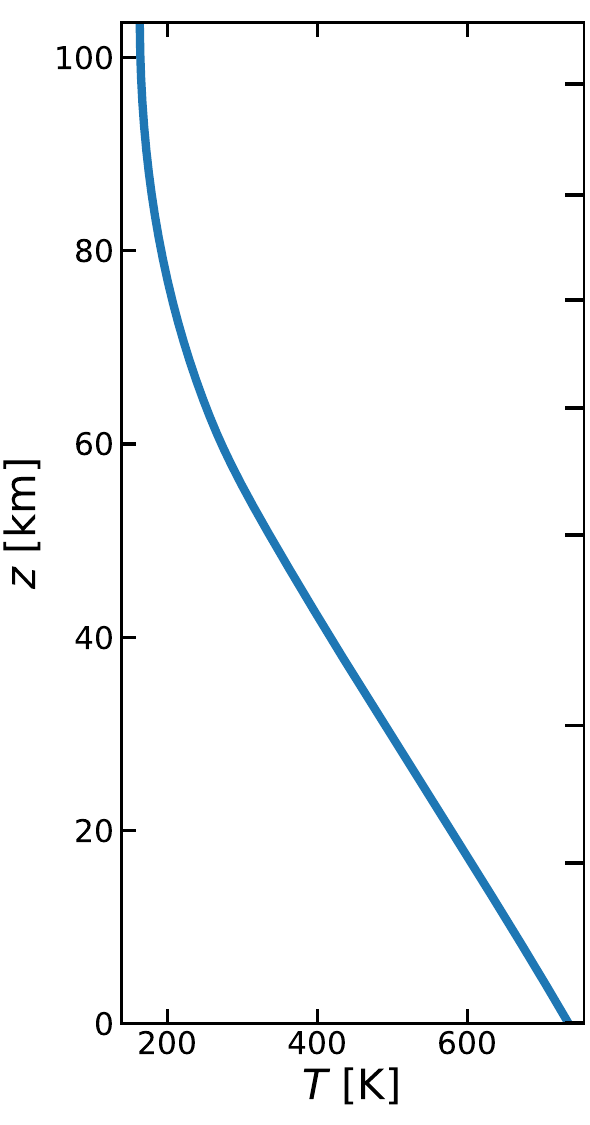} &
  \hspace*{-5mm}
  \includegraphics[page=6,height=100mm,trim=61 0 50 0,clip]
                  {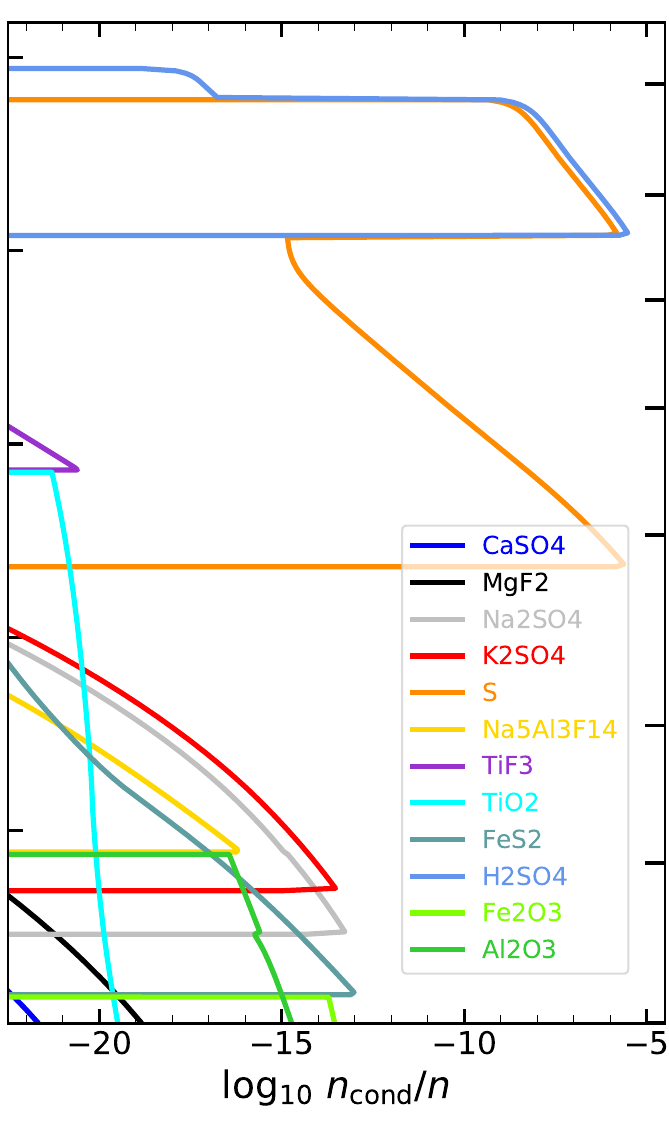} &
  \hspace*{-5mm}
  \includegraphics[page=5,height=100mm,trim=61 0 0 0,clip]
                  {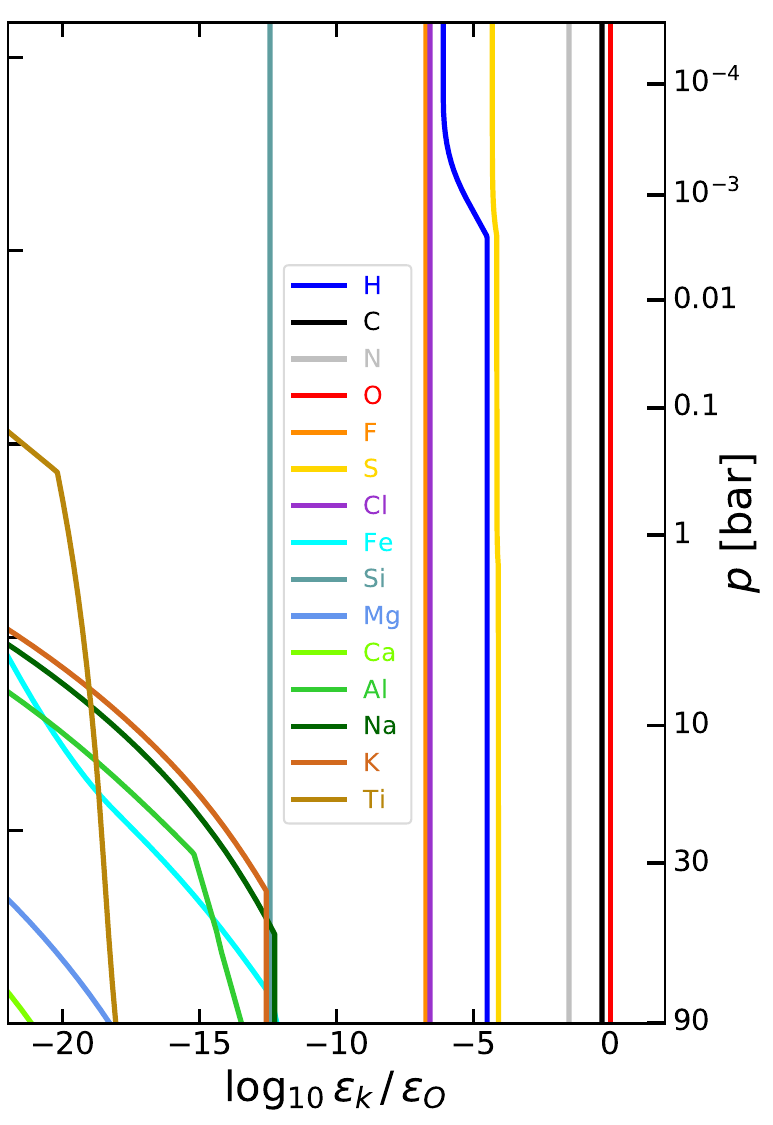}             
  \end{tabular}
  \vspace*{-4mm}
  \caption{{\bf Left:} Assumed temperature/pressure structure of the Venus
    atmosphere. {\bf Centre:} The concentrations of condensed units
    $n_{\rm cond}/n$ that deposit and are removed from the model when advancing to the
    next atmospheric height. {\bf Right:} Element abundances with
    respect to oxygen $\epsilon_k/\epsilon_{\rm O}$ remaining in the gas phase.}
  \label{fig:ggchem}
  \vspace*{2mm}
\end{figure*}

{In Fig.\,\ref{fig:ggchem} we show the results of a new, simple {\sc GGchem} model for the near-surface layers, where we seek to establish which condensates can be expected to form at which height in the Venus atmosphere.} 
We consider a simple 1D atmospheric structure with a given pressure 
and temperature profile.  The $(p,T)$-structure is obtained from a
spline fit to the values given by \cite{Palen2014}, using 1000
equidistant points between 0 and about 105\,km. At the bottom 
of this structure, our previous {\sc GGchem} model \citep{Rimmer2021}
applies.

The cloud structure of the Venus atmosphere is then calculated as
described by \citet{Herbort2022}.  In each atmospheric layer, {\sc
  GGchem} determines the stable condensates and the amount of elements
that are consumed to form these condensates, {until saturation is reached. The respective numbers of condensed units per volume $n_{\rm cond}$ are stored and the corresponding amounts of condensed elements are subtracted from the total element abundances $\epsilon_k$}, before advancing upwards to the next atmospheric layer.  Precisely speaking, $n_{\rm cond}$ is the number of
condensed units per $\rm cm^3$ that deposit per height interval
$\Delta z$, when a considered gas parcel is lifted by $\Delta z$
in the atmosphere.  Since the next layer is generally cooler, additional
condensation will take place and new cloud condensates may occur.  

{We note that we are using the assumption of chemical equilibrium in the gas phase while determining the stability of condensates in this model, which is a simplification. Chemical equilibrium is not expected to hold in the Venus atmosphere \citep{Zolotov2018,Zolotov2021}, except for the near-surface layers \citep{Zolotov1996, Krasnopolsky2013, Bierson2020, Rimmer2021}. However, the details of the gas phase chemistry have little influence on the principle stability of condensates in the atmosphere, because of the very steep dependences of the vapour pressures as function of temperature, and our method is in line with many theoretical models used in exoplanet science to establish the cloud structures, e.g. \citet{Allard2001}, \citet{Morley2012}, and \citet{Schaefer2012}.} 

We updated the {\sc GGchem} code with new Gibbs free energy data 
from FastChem \citep{Kitzmann2024}, based on the NIST/Janaf data
\citep{Chase1998} of 96 additional condensed species, in particular
sulphates, sulphides, carbonates, nitrides, hydroxides, hydrides,
fluorides, chlorides, carbides, and cyanides. {Thanks to the referee, we identified an error in interpreting the vapour pressures of di-sulphur \ce{S2} and octa-sulphur \ce{S8} as given by \cite{Zahnle2016}, see their equations (4) and (5).  These vapour pressures are given over solid and liquid sulphur \ce{S[s/l]}, not over the respective pure condensed phases as required in GGchem. It seems that \ce{S2} has no condensed phase at all, so we eliminated \ce{S2[s]} and \ce{S2[l]} from GGchem, and have now included only \ce{S8[s]} according to \cite{Kasting1989}, who give the vapour pressure of gaseous \ce{S8} over \ce{S8[s]} as required in GGchem. The alpha-phase of solid sulphur \ce{S[s]} {\sl (orthorhombic sulphur)} is not to be confused with solid octa-sulphur \ce{S8[s]}.}

The resulting cloud structure, $n_{\rm cond}(z)$, and the element
abundances remaining in the gas phase, $\epsilon_k(z)$, are shown
in Fig.\,\ref{fig:ggchem}. The main results are:
\begin{enumerate}
\setlength\parskip{-0.5em}
\setlength\itemsep{0.7em}
\item There are four groups of elements in the Venus atmosphere
  with abundance hierarchy $\rm O,C,N \gg S,H \gg Cl,F \gg Fe, 
  Na, Si, K, Al, Ti, Mg, Ca$. This hierarchy immediately suggests
  that the trace metals in the last group prefer to form chloride
  and fluoride molecules rather than oxides. 
\item Below the main sulphuric acid clouds $\la\!\!45$\,km, 
  called the ``lower thin haze'' by \cite{Knollenberg1980}, 
  {we find a sequence of three main layers of new stable condensates:} pyrite
  \ce{FeS2[s]} at $z\ga 2.9$\,km, sodium sulphate \ce{Na2SO4[s]}
  at $z\ga 9$\,km and potassium sulphate \ce{K2SO4[s]} at 
  $z\ga 14$\,km. As pointed out by \cite{Byrne2024}, the
  prediction of the stability of pyrite, {a conducting material,} about 3\,km above ground
  by our {\sc GGchem} model provides a natural explanation of 
  the strongly reflective mountain tops observed in radar
  measurements of the Pioneer Venus and Magellan missions
  \citep{Pettengill1982,Pettengill1996}, similar to the snow
  coverage of the mountain tops on Earth. {The stability of pyrite in the highlands was assessed earlier by, e.g., \cite{Barsukov1982}, \cite{Klose1992}, and \cite{Zolotov1991, Zolotov1992}.} The 
  formation of sulphate clouds (\ce{Na2SO4[s]} and \ce{K2SO4[s]}) 
  is a new finding.

\item In addition to these three condensates, the following solid
  materials are part of the surface and remain stable in the gas over
  the surface, but in less quantities: 
  \ce{Fe2O3[s]} {\sl (hematite)},
  \ce{CaSO4[s]} {\sl (anhydrite)},
  \ce{Al2O3[s]} {\sl (corundum)},
  \ce{TiO2[s]} {\sl (rutile)}, and
  \ce{MgF2[s]} {\sl (magnesium ﬂuoride)}.

\item The gas phase element abundances of all metals but Si (that is
  Fe, Na, K, Al, Ti, Mg, Ca) are subsequently reduced by orders of
  magnitude above their respective cloud bases via condensation and
  gravitational settling.
  %, i.e.\ above the heights above which a cloud
  %condensate is found to be thermally stable.
  
\item Silicon is found to stay entirely bound in Si-tetrafluoride
  \ce{SiF4} and does not participate in cloud condensation. 

\item Our {\sc GGchem} model does not work for the main sulphuric acid
  clouds, which only form at $z\!\ga\!80\,$km in this model. Instead,
  {we get crystalline sulphur, \ce{S[s]} at $z\!\ga\!45$\,km.}
  This is not unexpected, because the \ce{H2SO4} clouds are known
  to form from the photodissociation products of \ce{CO2} and
  \ce{SO2}, which is not included in our model.

\item Above the \ce{H2SO4[s]} cloud base, at $z\!\ga\!80\,$km in our
  {\sc GGchem} model, the hydrogen abundance falls rapidly, eventually
  approaching the sum of the chlorine and fluorine abundances, to
  retain the concentrations of HCl and HF, whereas the sulphur
  abundance only decreases very slightly.  This is because of the
  stoichiometry of \ce{H2SO4} and the relation $\rm S > H$ present at
  the bottom of the atmosphere. The formation of sulphuric acid clouds
  is hence limited by H rather than by S. This is in contrast to
  observations of \ce{H2O} and \ce{SO2} above the clouds and called
  the {\em puzzle of sulphur depletion} by \citet{Rimmer2021}.
\end{enumerate}

%-------------------------------------------------------------------
\section{The DiffuDrift v2 model}
%-------------------------------------------------------------------
\label{sec:DiffuDrift}
\noindent
While our {\sc GGchem} model discussed in Sect.~\ref{sec:GGchem} can
predict which cloud materials become thermally stable at which heights
in the Venus atmosphere, it cannot make quantitative predictions about
the amount of cloud particles nor their sizes.  For example, in a
completely quiescent atmosphere without any turbulent mixing, all
cloud condensates would eventually settle down, and no cloud particles
or droplets would remain.

In order to make predictions about the amount and sizes of the cloud
particles, a kinetic description of their growth and dynamical
behaviour is required.  The {\sc DiffuDrift} code was introduced by
\cite{Woitke2020} to model cloud formation in diffusive atmospheres of
brown dwarfs and exoplanets. The code simulates cloud particles in
terms of their size moments $L_j\;\{j\!=\!0,1,2,3\}$, similar to
\cite{Woitke2003,Woitke2004} and \cite{Helling2006},
\begin{equation}
  \rho L_j = \int_{V_\ell}^\infty f(V)\,V^{j/3}\,dV
\end{equation}
where $V\,\rm[cm^3]$ is the volume of a cloud particle and
$f(V)\,\rm[cm^{-6}]$ is the cloud particle size distribution
function. $\rho\rm\,[g/cm^3]$ is the gas mass density, and $V_\ell$ is
a lower integration boundary for particles to make sure they have
macroscopic properties.  The unit of $L_j$ is $[\rm cm^j/g]$,
for example $L_0$ is the number of cloud particles per gram of gas,
and $L_3$ is the total volume of the cloud particles per
gram of gas.

The modelling concept of {\sc DiffuDrift} is to evolve the cloud
particle moments in time, in a 1D hydrostatic atmosphere, as affected
by nucleation, growth~\&~evaporation, coagulation, gravitational
settling and diffusion
\begin{eqnarray}
  \frac{d(\rho L_j)}{dt} 
  &=&
      \underbrace{\Vl^{j/3} J_\star}_{\rm nucleation}
   + \underbrace{\int_{V_\ell}^{\infty}\!\frac{\partial V}{\partial t}
                  f(V)\,V^{j/3}\,dV}_{\rm growth\;\&\;evaporation}\,
   + \underbrace{\frac{\partial(\rho L_j)}{\partial t}\Bigg|_{\rm coag} 
                 }_{\rm coagulation}
  \nonumber\\
  &+& 
  \underbrace{\frac{\partial}{\partial z}
    \int_{V_\ell}^{\infty}\!\vdreq\,f(V)\,V^{j/3}\,dV}_{\rm settling}  
  +\underbrace{\frac{\partial}{\partial z}\,
      \bigg(D\,\rho\,\frac{\partial L_j}{\partial z}\bigg)}_{\rm diffusion} 
  \,,\label{eq:mom}
\end{eqnarray}
where $J_\star\rm\,[s^{-1}cm^{-3}]$ is the nucleation rate, i.e.\ the
formation rate of seed particles of size $V_\ell$ that form via
chemical processes in the gas phase, $\partial V/\partial t$ is the
change of a particle's volume due to the deposition and sublimation of
molecules on its surface (see App.\,\ref{AppB}), $\partial(\rho
L_j)/\partial t\,|_{\rm coag}$ is the change of the cloud particle
moments due to particle-particle collisions, $\vdreq\rm\,[cm/s]$ is
the downward equilibrium drift velocity (or terminal fall speed), and
$D\rm\,[cm^2/s]$ is the eddy-diffusion coefficient due to turbulent
mixing in the atmosphere.

The advantage of this modelling concept is that, for example in the
Epstein regime, $\partial V/\partial t\propto a^2\rho$ and
$\vdreq\!\propto\!a/\rho$ depend on particle radius
$a\!=(3V/(4\pi))^{1/3}$ with certain powerlaws, so that the
two integrals over particle volume in Eq.\,(\ref{eq:mom}) can be
solved analytically, leading to a term involving $L_{j-1}$ for the net
growth, and a term involving $L_{j+1}$ for the settling
\citep{Woitke2003}.  The size-distribution function $f(V,z)$ is not a
direct result of the {\sc DiffuDrift} model, but all relevant mean
particle properties, such as the mean size, total surface, etc., can
be calculated from the cloud particle moments $L_j(z)$, which has the
advantage that we do not need to solve hundreds of equations for
pre-defined size bins, but can approximately solve the problem of
cloud formation with only four equations.

However, in the previous version of {\sc DiffuDrift}
\citep{Woitke2020}, we explicitly used the special case of a subsonic
flow of the gas around the particles and large Knudsen numbers
(Epstein regime), but in dense atmospheres like the ones of Earth and
Venus, or in the deeper layers of gas giants including hot Jupiters,
this is incorrect.  We have therefore extended our moment method to
arbitrary Knudsen numbers in this paper, see App.~\ref{AppA} and
\ref{AppB}.  Furthermore, coagulation caused by (i) Brownian motion,
(ii) difference fall speeds, and (iii) turbulence has been included,
see App.~\ref{AppC}.

The key idea for arbitrary Knudsen numbers is to use
a double-$\delta$ representation of the particle size distribution
function $f(V)$, inspired by \cite{Birnstiel2012}, as
\begin{equation}
  \rho L_j = \sum_{i=1}^2 n_i V_i^{\,j/3} \quad\quad(j=0,1,2,3) .
  \label{eq:double-d}
\end{equation}
$n_1$ and $n_2$ are two representative particle densities, and $V_1$
and $V_2$ their respective volumes. The mapping $\{L_0,L_1,L_2,L_3\}
\leftrightarrow \{n_1,V_1,n_2,V_2\}$ is unique in a sense that any
given set of $\{n_1,V_1,n_2,V_2\}$ results in a particular set of
moments and vice versa.  In \cite{Woitke2020}, this approximation
(Eq.\,\ref{eq:double-d}) was used anyway, as closure condition, to
construct $L_4$ from $\{L_0,L_1,L_2,L_3\}$.  Here, we use it prior to
the calculations of any of the r.h.s.\ terms in Eq.\,(\ref{eq:mom}) at
any height, in which case a closure condition is no more necessary.

In App.~\ref{AppA} and \ref{AppB} we show that
Eq.\,(\ref{eq:double-d}) becomes an exact representation of the
size distribution when all particles are either in
the Epstein or in the viscous Stokes regime.  In both Knudsen number
limiting cases, the respective terms for growth and
settling can be computed either by (i) the integrals on the r.h.s.\ of
Eq.(\ref{eq:mom}) or by (ii) applying the double-$\delta$
representation (Eq.\,\ref{eq:double-d}) first and then calculate the
terms based on the properties of the two representative particles, as
formulated on the r.h.s.\ of Eq.\,(\ref{eq:Lj}) -- the results are
identical. Only when one of the two representative particles is in one
and the other particle in another hydrodynamical regime, using
Eq.\,(\ref{eq:double-d}) becomes an approximation.

As explained by \citet{Woitke2020}, {\sc DiffuDrift} considers a
mixture of condensed materials $V\!=\!\sum_s V^s$, where $s$ is an
index for the included solid and liquid materials. We assume that all
particles of any size at one point in the model are made of the same
material mixture, however this mixture can change with time and space.
The book-keeping of condensates works by introducing one additional
equation per material, using the volumes of each kind $L_3\!=\!\sum_s
L_3^s$. Furthermore, {\sc DiffuDrift} keeps track of the element
abundances in the gas phase, accounting for the consumption and
enrichment due to nucleation, particle growth and evaporation.  The
complete set of modelling equations is
\begin{eqnarray}
  \frac{d(\rho L_j)}{dt} 
  &\,=\,& \Vl^{j/3} J_\star
  \,+\, \frac{j}{3} \sum_{i=1}^2 n_i V_i^{\;j/3\,-1}\,\frac{\partial V_i}{\partial t}
  \,+\, \frac{\partial(\rho L_j)}{\partial t}\Bigg|_{\rm coag} 
    \nonumber\\[-2mm]
  &+&   \frac{\partial}{\partial z} \sum_{i=1}^2 n_i V_i^{\;j/3}\,\vdreq_{,i}
  \,+\, \frac{\partial}{\partial z}\,
        \bigg(D\,\rho\,\frac{\partial L_j}{\partial z}\bigg)
  \label{eq:Lj}
\end{eqnarray}
\vspace*{-4mm}
\begin{eqnarray}      
  \frac{d(\rho L_3^s)}{dt} 
  &\,=\,& \Vl\,J_\star^s
  \,+\, \sum_{i=1}^2 n_i\,\frac{\partial V_i^s}{\partial t}
        \nonumber\\[-2mm]
  &+&   \frac{\partial}{\partial z} \sum_{i=1}^2 n_i V_i^s\,\vdreq_{,i}
  \,+\, \frac{\partial}{\partial z}\,
        \bigg(D\,\rho\,\frac{\partial L_3^s}{\partial z}\bigg)
  \label{eq:L3s}
\end{eqnarray}
\vspace*{-4mm}
\begin{eqnarray}
  \frac{d(\rho \epsilon_k)}{dt} 
  &\,=\,& -\sum_s \frac{\nu_{s,k}\Vl}{V_0^s} J_\star^s
  \,-\, \sum_s \frac{\nu_{s,k}}{V_0^s} \sum_{i=1}^2 n_i\,\frac{\partial
        V_i^s}{\partial t}
    \nonumber\\[-1mm]
  &+& \frac{\partial}{\partial z}\,
      \bigg(D\,\rho\,\frac{\partial \epsilon_k}{\partial z}\bigg) \,,
  \label{eq:epsk}
\end{eqnarray}
where $\{j\!=\!0,1,2,3\}$ is the moment index, $\{s\!=\!1,...,S\}$ the
index for the included condensates, and $\{k\!=\!1,...,K\}$ the index
of the elements affected by condensation.
As can easily be verified, $\sum_s$(Eq.\,\ref{eq:L3s}) is identical to
Eq.\,(\ref{eq:Lj}) with $j\!=\!3$.  $V_i\!=\!\sum_s V_i^s$ is the total
volume of the representative particle $i$ which is composed of
the volumes of the different materials $V_i^s$. $\vdreq_{,i}$ is the
fall speed of the representative particle $i$.  $\epsilon_k$ is the
abundance of element $k$, i.e. the number of nuclei of element $k$ per
gram of gas, $J_\star^s$ is the nucleation rate of material
$s$. $V_0^s$ is the volume of one condensed unit in material
$s$ and $\nu_{s,k}$ is the stoichiometric coefficient of element $k$ in
material $s$. Coagulation does not change the total volume of
condensates, nor does it have any influence on the gas; hence there
are no terms for coagulation in Eqs.\,(\ref{eq:L3s}) and
(\ref{eq:epsk}).

\begin{figure*}
  \includegraphics[width=180mm]{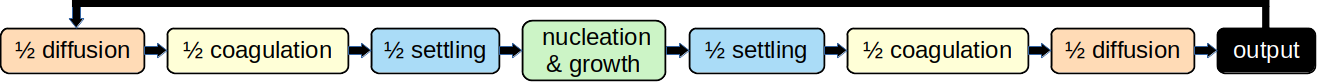}
  \caption{Numerical method to simulate the various physical and
    chemical processes one by one. The notation $1/2$ means to apply
    half a time step. The nested sequence of the operators assures
    $2^{\rm nd}$ order accuracy in time, as known from fluid dynamics
    simulations \citep[e.g.][]{Woodward1984}.}
  \label{fig:OPsplit}
  \vspace*{2mm}
\end{figure*}

\subsection{Gas phase chemistry}
\label{fig:chemistry}
\noindent
To determine the local net growth rates $\partial V/\partial t
= \sum_s \partial V_i^s/\partial t$, we assume chemical equilibrium in
the gas phase and call {\sc GGchem} at each time step and on each
atmospheric grid point as
\begin{equation}
  n_{\rm mol} = n_{\rm mol}(\,\rho,T,\epsilon_k) \ .
\end{equation}
This application of {\sc GGchem} is for the gas phase only,
disregarding condensation, to determine all molecular particle
densities $n_{\rm mol}$ and the supersaturation ratios $S$ of all
considered cloud condensates. {We again note that this is an approximation as we cannot account for disequilibrium effects like photodissociation this way}. These supersaturation ratios are then
used to determine the volume growth rates of all materials $s$ for the
two representative particle sizes $i$, $\partial V_i^s/\partial t$.
Equation~(12) in \cite{Woitke2020} shows that these growth rates, in
the case of large Knudsen numbers, scale with $\,n_r^{\rm key}(1-1/S)$,
where $n_r^{\rm key}$ is the particle density of the least abundant
reactant (the key species) in a considered surface reaction $r$.  For
small Knudsen numbers, the total growth rate is given by Eq.\,(32) in
\cite{Woitke2003}, which also scales with $\,n_r^{\rm key}(1-1/S)$, and
can be generalised to a mixture of different materials in the same
way. Further details about the growth rates are in
App.~\ref{AppB}.  In all hypothetical surface reactions listed in
Table~\ref{tab:conc}, the first reactant (a metal chloride of metal
fluoride molecule) is by far the least abundant reactant and hence the
growth rates are determined by the abundances of these rare molecules.

\subsection{Coagulation}
\label{fig:coag}
\noindent
A new coagulation module has been added to {\sc DiffuDrift v2}, based
on a numerical implementation of the Smoluchowski equation
\citep{Smoluchowski1916}, taking into account particle collisions due
to Brownian motions, turbulence, and different settling velocities of
the two colliding particles, see App.~\ref{AppC}. We have also
implemented the repelling effect of particle charges, using an
electrostatic repulsion factor $f^{\rm C}_{i,j} = \exp\big(-E_{\rm
  C}/E_{\rm kin}\big)$ as explained in App.\,\ref{AppC}, see
Eq.\,(\ref{eq:frep}), where $E_{\rm C} = q_i\,q_j\,{\rm
  e}^2/(a_i+a_j)$ is the Coulomb energy, $q_i$ and $q_j$ are the
numbers of negative charges on the colliding particles $i$ and $j$,
and $a_i$ and $a_j$ are the radii of the colliding particles.
Concerning the particle charges, we use the scaling behaviour found by
\cite{Balduin2023} in UV-shielded environments, also explained in the
Appendix
\begin{equation}
  q = -\,\mbox{qa300} \left(\frac{a}{1\rm\,\mu m}\right)
                       \left(\frac{T}{300\rm\,K}\right) \ ,
  \label{eq:qa300}
\end{equation}
where $a$ is the particle radius and qa300 is a numerical parameter, the
number of negative charges on a 1\,$\mu$m particle at $T=300\,$K,
henceforth called the 'grain charge parameter', that is varied from
a few to a few tens in this paper. {As shown by \citet{Balduin2023},
even a very small but non-zero cosmic ray ionisation rate creates pairs of free electrons and molecular cations. Since the electrons move faster, they charge the particles negatively, and the effect on the particles scales with $q/a/T$. Therefore, particles of all sizes in one location obtain similar $q/a$ ratios.}

\subsection{Operator splitting method}
\noindent
Figure~\ref{fig:OPsplit} shows the numerical method used to advance
our {\sc DiffuDrift v2} model in time.  Each block represents an
operator, which advances a single source term on the right sides of
Eqs.\,(\ref{eq:Lj}) to (\ref{eq:epsk}) in time. For numerical
stability it is important that each operator changes the
physico-chemical state only moderately at any atmospheric layer,
i.e.\ not far from it's linear behaviour.  For example, the nucleation
\& growth must not entirely consume all condensible elements in one
time step, in other words, the element abundances must not change
significantly during $\Delta t$.  Concerning the settling, the time
step $\Delta t$ must remain small enough to ensure that $\vdreq_{,i}\,\Delta
t\!<\!\alpha \Delta z$, where $\Delta z$ is the vertical grid
resolution.  Similar criteria apply to the coagulation and diffusion
operators. 
We choose a precision of $\alpha\!=\!0.3$ for the models presented in
this paper, which results in typical $\Delta t$ of about $100-1000$\,s,
while we need to advance the atmosphere by a least a few 100 days to
reach the stationary limit. The different operators are characterised
by:
\begin{itemize}
\item The diffusion is solved by a standard $2^{\rm nd}$ order explicit
  numerical scheme.  The lower and upper boundaries of all variables
  $\{L_j,L_3^s,\epsilon_k\}$ can be individually chosen to be either
  constant concentration or constant influx/outflux.
\item The coagulation is computed in an explicit way as explained
  in App.~\ref{AppC}. 
\item The settling is simulated optionally by a $1^{\rm st}$ or a new
  $2^{\rm nd}$ order explicit upwind scheme.
\item The nucleation and growth is advanced in time by a new fully
  implicit integration scheme.  
\item We have added a new module for the calculation of particle
  opacities based on effective mixing and Mie theory, only used
  for output, see Sect.\,\ref{sec:kappa}.  
\end{itemize}
More details about the numerical methods used in {\sc DiffuDrift} can
be found in the appendix of \cite{Woitke2020}.

%What is new in {\sc DiffuDrift v2} is that we have included (i)
%coagulation of charged particles, (ii) a general treatment of growth
%and settling for arbitrary Knudsen numbers, (iii) an option for
%$2^{\rm nd}$ order settling, and (iv) a new fully implicit integration
%method for the stiff problem of nucleation and growth instead of
%calling an ODE solver, which results in a large computational
%acceleration.

\subsection{Particle opacities}
\label{sec:kappa}
\noindent
The extinction opacities of the particles are calculated via the
representative particles, which together have the correct total cross
section and mass
\begin{equation}
  \kappa^{\rm ext}_\lambda ~=~
       \sum\limits_{i=2}^2 n_i\,\pi a_i^2\,Q_{\rm ext}(a_i,\lambda) \ .
  %\tau_\lambda(z) &=& \int_z^\infty \!\!\kappa^{\rm ext}_\lambda(z')\,dz' 
  %\quad\mbox{,}\quad {\rm visibility} ~=~ 1/\kappa^{\rm ext}_\lambda \ .
  \label{eq:visib}
\end{equation}
The extinction efficiencies $Q_{\rm ext}(a,\lambda)$ are calculated
with effective mixing and Mie theory, based on the particle radius
$a$, the wavelength $\lambda$, and the local volume composition of the
particles $\{L_3^s/L_3|\,s=1,...,S\}$.  The details are explained in
\cite{Woitke2024}, where the references for our optical data
(refractory indices) are given in table A.1.  Unfortunately, out of
the six materials considered in this paper, only \ce{TiO2} and
\ce{Al2O3} have valid optical data. For \ce{FeS2}, we use the optical
data for \ce{FeS} (troilite) instead. Both materials are conducting
and are hence very opaque in the optical.  For the sulphates
\ce{K2SO4}, \ce{Na2SO4} and \ce{CaSO4}, we use the optical data of
\ce{Na2S} instead, and for \ce{MgF2}, we use the optical data of
\ce{MgO}.  Most relevant, for the ``passive'' particles, we use the
optical data of amorphous Mg$_{0.7}$Fe$_{0.3}$SiO$_3$, a common
pyroxene. At optical wavelengths, for sub-micron particles, we find
extinction efficiencies $Q_{\rm ext} \approx 1.3-2.1$ depending on the
choice of material, which reflects our uncertainty in the optical
data.

\subsection{The atmospheric structure}
\noindent
All models for the lower Venus atmosphere presented in this paper are
based on the $(p,T)$-structure based on \cite{Palen2014} as introduced
in Sect.\,\ref{sec:GGchem}.  The eddy diffusion constant is assumed to be
\begin{equation}
  D(z) = 3\times 10^4\;{\rm\!cm^2 s^{-1}}\;\bigg(\frac{p}{1\rm\,bar}\bigg)^{-1/2}
  \label{eq:eddy_diff} \ ,
\end{equation}
which is a smooth version of $D(z)$ as used by \cite{Rimmer2021}.

\subsection{Boundary conditions}
\label{sec:BC}
\noindent
Solving the diffusion problem requires to set boundary conditions
(BCs) at the lower and upper model domain. The {\sc DiffuDrift v2}
code offers various options for the choice of the BCs.  For our models in
this paper, we have selected fixed-concentration BCs for the gas-phase
element abundances $\epsilon_k$ and the moments of the passive
particles $L_j$ at the lower boundary. For the material composition of
these particles $L_3^s$ we choose zero-concentration lower BCs for the
thermally unstable materials and zero-flux lower BCs for the thermally
stable materials (\ce{Al2O3}, \ce{TiO2}, \ce{MgF2} and
\ce{CaSO4}). Concerning the upper boundary, we use zero-flux BCs for
all components of the solution vector $\{L_j,L_3^s,\epsilon_k\}$.
%In a forthcoming paper, we will explore to use constant-influx BCs at
%the upper boundary to simulate the influx of particles (or gas-phase
%elements) via micro-meteorites.

The lower BC fixed-concentration values for the element abundances
$\epsilon_k$ are those predicted by our GGchem model at $z=0$ (see
Sect.\,\ref{sec:GGchem}). In order to set the fixed values for
$L_j$ at the lower boundary, we need to make assumptions about the
total number density of the particles and their size distribution
$f(a)$ at $z=0$.

From the Pioneer Venus sounder probe data, \cite{Knollenberg1980}
estimated a mean particle diameter of 0.25\,$\mu$m, i.e.\,$\langle
a\rangle=0.125\,\mu$m (mode-1 particles) at $z=40\,$km, although the
instruments were only designed to detect particles with a diameter
larger than $0.6\,\mu$m. They arrived at this conclusion by fitting
their size-dependent data with a log-normal distribution as visualised
by \cite{Seager2021}, see their Fig.\,2. Based on the discharge
current measurements by Venera 13 and 14, \cite{Lorenz2018} estimated
a total charge density of about 1500\,pC/m$^3$ at $z=0$. From
that value, they estimated an optical extinction opacity of 0.5/km at
$z=0$, assuming the particles to be about singly charged.  Based on
the spectrophotometer data on Venera 13 and 14, \cite{Grieger2004}
found opacities of order 1/km, which drop to about 0.1/km at a height
of 40\,km.
%The results of this paper will show that coagulation should be very
%efficient to grow and rain-out all particles quickly, if the particle
%charges are indeed that small, which is not observed.  We therefore
%use the measured charge/volume value, rather than the opacity values,
%to set our lower BC.

Based on these observations, we use a log-normal distribution
$\rm [particles/cm^3/\mu m]$ defined as
\begin{equation}
  f(a,z\!=\!0) \,=\, \frac{n_p}{a\,\sigma\sqrt{2\pi}}
         \exp\bigg(-\frac{\big(\ln a-\mu\big)^2}{2\sigma^2}\bigg) 
\end{equation}
for our lower BC with $n_p=5000\rm\,cm^{-3}$, $\mu=\ln(0.15)$ and
$\sigma=0.5$, using 50 size-bins between $0.005\,\mu$m and $20\,\mu$m
to make sure that the particles outside of this size range can be
neglected. Assuming a grain charge parameter of $\rm qa300=4$, this
results in a charge density of 1540\,pC/m$^3$ and an opacity of 1.8/km
at $z=0$.

%tauext,tauabs,kext[1/km],change[pC/m3] =
%2.458E+01  4.906E-01 1.821E+00  1.520E+03

Our opacity model shows that the aerosol opacity is mostly scattering
opacity. The optical absorption opacity is only 0.089/km at $z=0$,
i.e.\ the albedo is about 95\%.  This is essential for the radiative
transfer in the Venus atmosphere.  For example, our model for passive
aerosol particles discussed in Sect.\,\ref{sec:passive} has a vertical
extinction optical depth of $\sim 25$, but the vertical absorption
optical depth is only $\sim 0.49$, still allowing for some sun light
to reach the Venus surface in a highly diffusive manner. To emphasise
the uncertainties in the opacities, we note that the surface pictures
from the Venera probes \citep{Florensky1977} show a clear horizon,
which suggests that the Venus atmosphere must be relatively
transparent at ground level, barely consistent with such large opacity
values, as also noticed by \cite{Sagan1975}.

\begin{figure}
  \hspace*{-2mm}
  \includegraphics[page=3,width=87mm,trim=15 15 10 10,clip]
                  {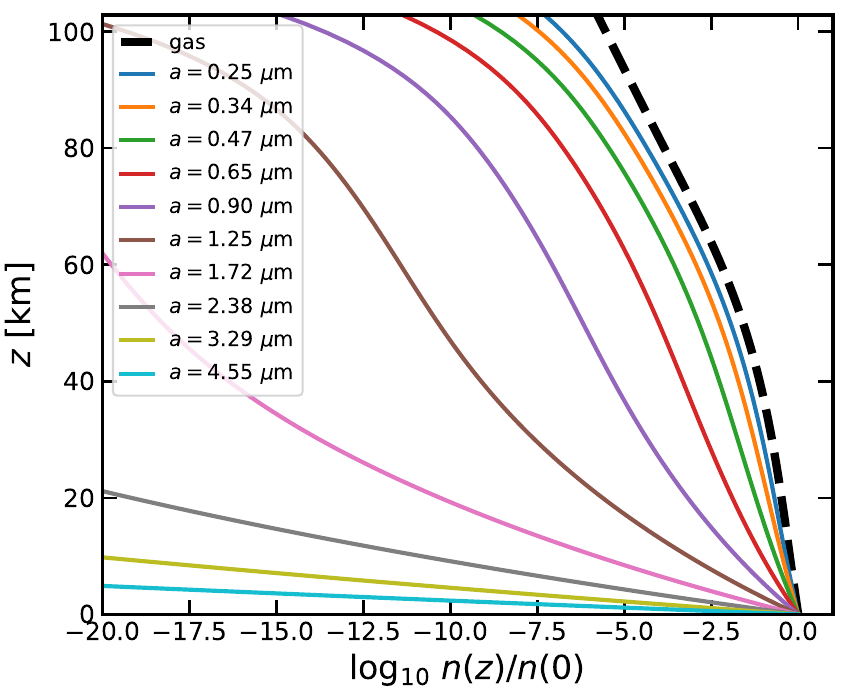}\\*[-6mm]
  \caption{The density of passive particles $n_p(z)$ in the Venus
    atmosphere with respect to their density over the surface $n_p(0)$
    as function of particle radius $a$. The eddy diffusion coefficient
    $D(z)$ is assumed to be given by Eq.\,(\ref{eq:eddy_diff}) and the
    material density is selected to be $\rho_{\rm m}\!=\!1.83\rm\,g/cm^3$ for this plot. The latter value corresponds to liquid \ce{H2SO4}, relevant for the uppermost regions in the plot.  Solid particles are expected to be porous, which causes the effective mass density to be substantially smaller than in a pure, e.g.\ basaltic material with $\rho_{\rm m}\!\approx\!2.9\rm\,g/cm^3$.}
  \label{fig:n_n0}
\end{figure}

\begin{figure*}
  %Series_03/Venus_passive/output/weather_latest.dat
  % 1        ! dust_diffuse
  % .false.  ! do_coagulation
  % 0.1      ! precision
  % .true.   ! upwind_2nd_order
  % 101      ! Npoints
  % 1.5      ! zpot
  % 2        ! bc_high
  % .false.  ! detailed_charges
  % 4.0      ! q_over_a
  % .true.   ! power_correct
  \vspace*{-2mm}
  \centering
  \begin{tabular}{cc}
  \hspace*{-5mm}
  \includegraphics[page=8 ,height=69mm,width=83mm,trim=0 0 0 0,clip]
                  {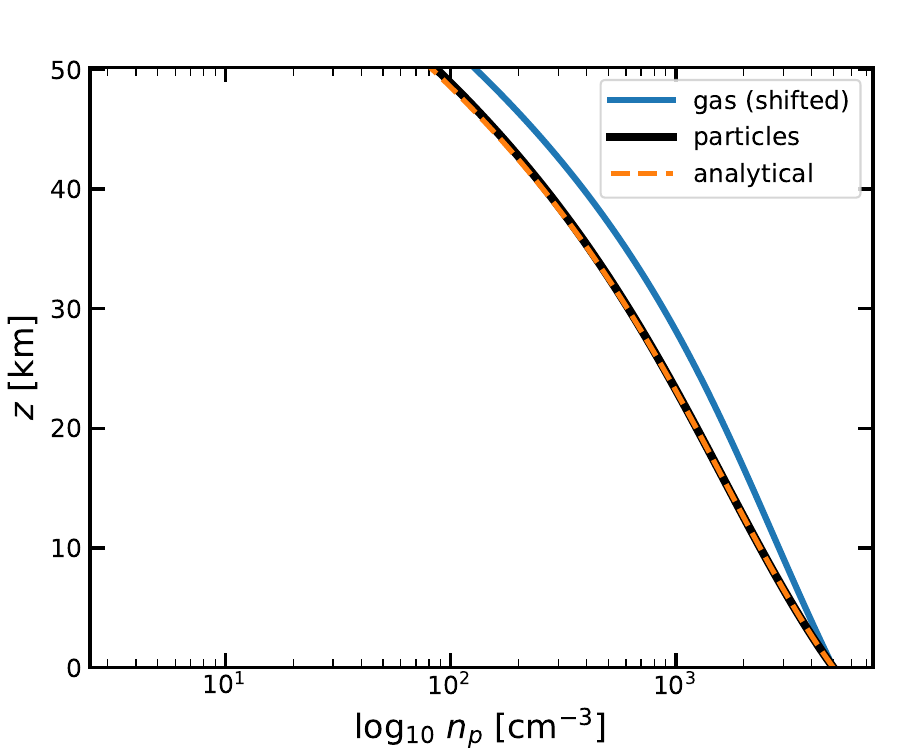} &
  \hspace*{-5mm}
  \includegraphics[page=7 ,height=69mm,width=83mm,trim=0 0 0 0,clip]
                  {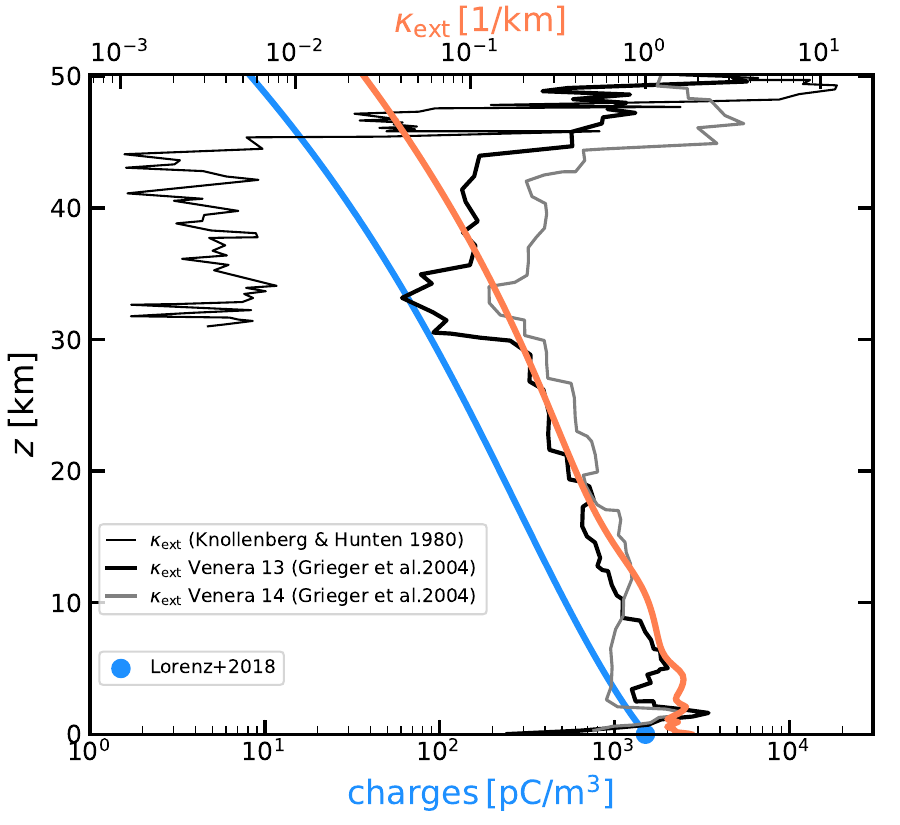} \\[-1mm]
  \hspace*{-5mm}
  \includegraphics[page=10,height=65mm,width=83mm,trim=0 0 0 25,clip]
                  {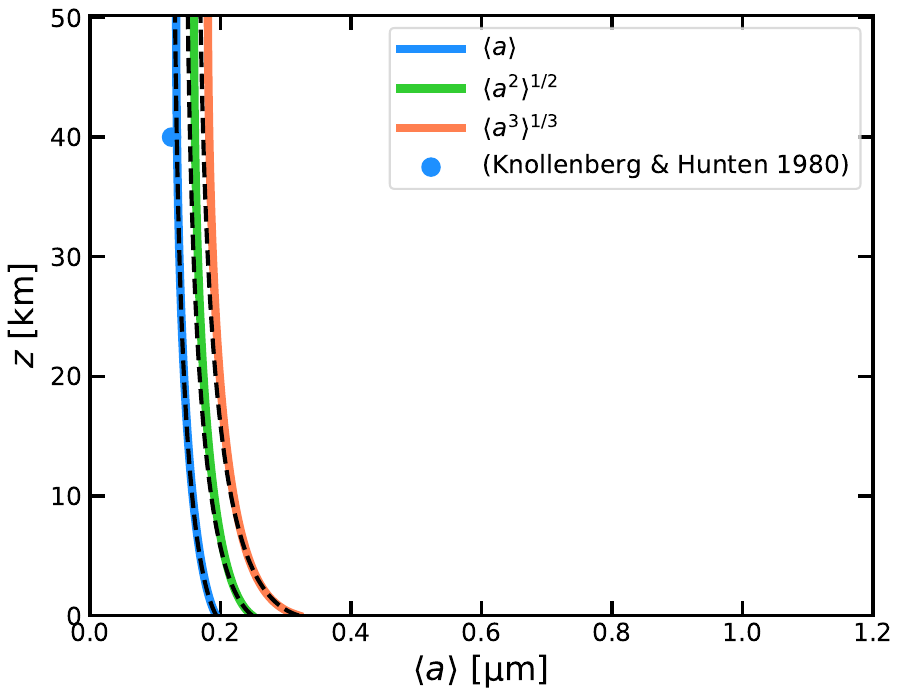} &
  \hspace*{-5mm}
  \includegraphics[page=11,height=65mm,width=83mm,trim=0 0 0 25,clip]
                  {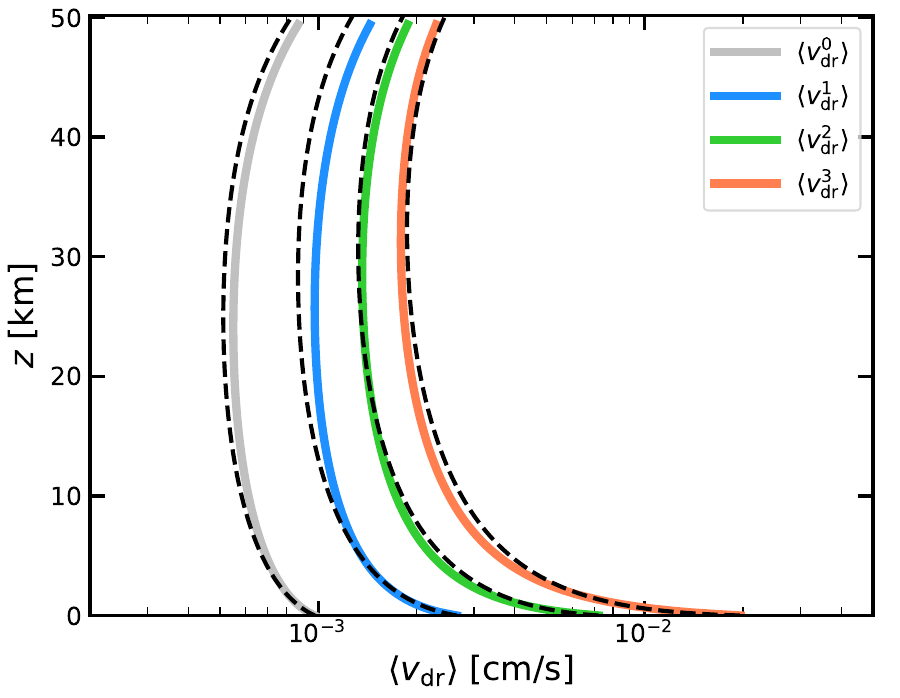} \\[-1mm]
  \end{tabular}
  \caption{Passive particles in the lower Venus atmosphere without
    coagulation, for grain charge parameter $\rm qa300=4$. The dashed
    lines represent the analytical solutions that we obtained from
    Eq.\,(\ref{eq:passive}). The bullets and additional black lines
    show selected measurements as discussed in the text. The data from
    \cite{Knollenberg1980} only includes the expected opacity of the
    detected particles with a diameter $\ga 0.6\,\mu$m ($a\ga
    0.3\,\mu$m). $\kappa_{\rm ext}$ is the extinction coefficient of the
    particles at $\lambda=600\,$nm.}
  \label{fig:passive_no}
  \vspace*{2mm}
\end{figure*}

%-------------------------------------------------------------------
\section{Results}
%-------------------------------------------------------------------
\label{sec:results}
\noindent
\subsection{Passive particles}
\label{sec:passive}
\noindent
Before we discuss our results for chemically active particles in the
lower Venus atmosphere, we first study the case of passive particles,
where we only have settling, diffusion and coagulation. This problem
has an analytical solution when we neglect coagulation, and this
solution can be used to test the {\sc DiffuDrift} implementation.
Considering one single particle size, the particle density
$n_p\!=\!\rho L_0\rm\,[cm^{-3}]$ follows from Eq.\,(\ref{eq:mom}) with
$j\!=\!0$
\begin{equation}
  \frac{dn_p}{dt} = \frac{\partial}{\partial z}\,\left(\vdreq\,n_p\right)
  ~+~\frac{\partial}{\partial z}\,\left(D\,\rho\,
     \frac{\partial}{\partial z}\bigg(\frac{n_p}{\rho}\bigg)\,\right)
  \ .
\end{equation}
Hence, in steady state $(dn_p/dt\!=\!0)$, there must be a constant
particle flux in time and space through the atmosphere
\begin{equation}
  \vdreq\,n_p
  ~+~ D\,\rho\,\frac{\partial}{\partial z}\bigg(\frac{n_p}{\rho}\bigg)
  ~=~ {\rm const} \ .
\end{equation}
Assuming that there is no influx nor outflux at the upper boundary
(neglecting the entry of micro-meteorites), the constant
is zero and we find
\begin{equation}
  \frac{\partial}{\partial z}\bigg(\ln\frac{n_p}{\rho}\bigg)
  ~=~ -\frac{\vdreq}{D} \ .
  \label{eq:passive}
\end{equation}
Equation (\ref{eq:passive}) means that the particle-to-gas ratio is
constant throughout the atmosphere when it is well-mixed ($D\gg\vdreq
H_p$), where $H_p$ is the atmospheric scale height. Otherwise, however,
the scale height of the particles is reduced, and $n_p/\rho$ drops
quickly with height, by orders of magnitude.

Equation (\ref{eq:passive}) can be integrated numerically, from the
bottom of the atmosphere upwards, to find $n_p(z)/n_p(0)$. These
solutions are shown for different particle sizes in
Fig.\,\ref{fig:n_n0}.  Particles smaller than about $a=0.2\,\mu$m are
expected to be well-mixed up to heights of about 100\,km in the Venus
atmosphere, but particles that are just slightly larger, $a\ga
1\,\mu$m, remain much more concentrated towards the bottom of the
atmosphere, barely reaching a height of 10\,km.  This is because of
the $\vdreq\!\propto\!a^2$ scaling in the Stokes regime as discussed
in App.\,\ref{AppA3}, which together with Eq.\,(\ref{eq:passive})
states that the scale height reduction is $100\times$ larger for
$10\times$ larger particles. Therefore, we expect a steep cutoff of
passive particles in size space around $1\,\mu$m at altitudes $\ga
10\,$km.

\begin{figure*}
  % Series_03/Venus_passive_coag
  % 1        ! dust_diffuse
  % .true.   ! do_coagulation
  % 2        ! coagulation_Nbin
  % 0.1      ! precision
  % .false.  ! upwind_2nd_order
  % 101      ! Npoints
  % 1.5      ! zpot
  % 2        ! bc_high
  % .false.  ! detailed_charges
  % ...      ! q_over_a
  % .true.   ! power_correct
  \centering
  \begin{tabular}{cc}
  \hspace*{-5mm}
  \includegraphics[page=2,height=61mm,width=80mm,trim=0 0 0 0,clip]
                  {multi_passive_yes.pdf} &
  \hspace*{-5mm}
  \includegraphics[page=1,height=61mm,width=80mm,trim=0 0 0 0,clip]
                  {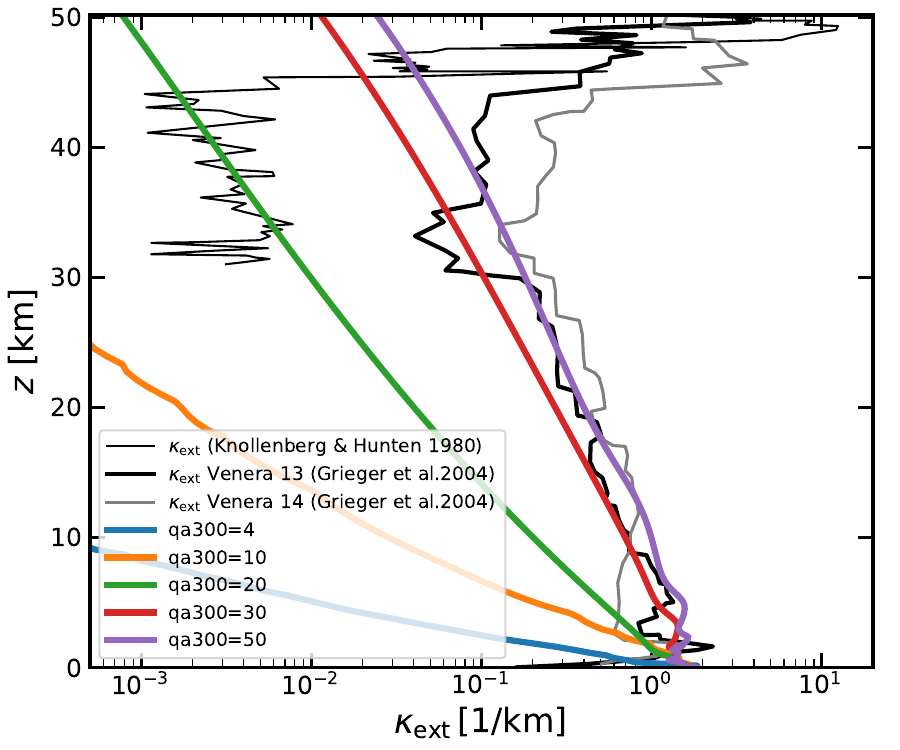} \\[-1mm]
  \hspace*{-5mm}
  \includegraphics[page=3,height=61mm,width=80mm,trim=0 0 0 0,clip]
                  {multi_passive_yes.pdf} &
  \hspace*{-5mm}
  \includegraphics[page=4,height=61mm,width=80mm,trim=0 0 0 0,clip]
                  {multi_passive_yes.pdf} \\[-2mm]
  \end{tabular}
  \caption{Same as Fig.\,\ref{fig:passive_no} but including
    coagulation for different values of the grain charge parameter
    qa300, see Eq.\,(\ref{eq:qa300}).}
  \label{fig:passive_yes}
  \vspace*{3mm}
\end{figure*}

Figure \ref{fig:passive_no} shows the results of a simple {\sc
  DiffuDrift v2} test model for chemically inert passive particles (no
growth and no coagulation) in the lower Venus atmosphere up to 50\,km,
where we assume the main sulphuric acid clouds to begin. The numerical
{\sc DiffuDrift} results are in good agreement with the semi-analytic
solutions obtained from Eq.\,(\ref{eq:passive}) shown with dashed
lines. The measurement values of the charge density at $z=0$ from
\cite{Lorenz2018}, and the mean particle radius $\langle a\rangle$ at
$z=40\,$km from \cite{Knollenberg1980} are shown with blue
bullets. In addition, the extinction opacity data from
\cite{Grieger2004} and \cite{Knollenberg1980} are shown with black and
grey lines. The average particle sizes and drift
velocities shown in Fig.\,\ref{fig:passive_no} are calculated as\\*[-1mm]
\begin{equation}
  \hspace*{-3mm}\langle a^j\rangle^{1/j} \,=\,
  \left(\frac{\int\limits_{V_\ell}^{\infty}\!a^j\,f(V)\,dV}
       {\int\limits_{V_\ell}^{\infty}\!f(V)\,dV}\right)^{1/j} \hspace*{-4mm}
  =\,\bigg(\frac{3}{4\pi}\bigg)^{1/3}\!\!\left(\frac{L_j}{L_0}\right)^{1/j}
  \!\!\!     
\end{equation}  
\begin{equation}
  \hspace*{-3mm}\langle \vdreq^{\!\!j}\rangle \,=\,
  \frac{\int\limits_{V_\ell}^{\infty}\!\vdreq\,f(V)\,V^{j/3}\,dV}
       {\int\limits_{V_\ell}^{\infty}\!f(V)\,V^{j/3}\,dV}
  \,\approx\,
  \frac{\sum\limits_{i=2}^2 \vdreq_{,i}\,n_i\,V_i^{j/3}}
       {\sum\limits_{i=2}^2 n_i\,V_i^{j/3}} \,.
  \label{eq:meanvd}
\end{equation}
Our simple model obtains a decent fit to all these data,
including the opacity slope in the lower Venus atmosphere. The wiggles
on $\kappa^{\rm ext}(z)$ are because of Mie-resonances that occur
around size parameters $x=2\pi\,a/\lambda\ga 1$, as the sizes of the
two representative particles change quickly near the surface, which
likely is an artefact of using only two particle sizes to calculate
the opacities.

Figure \ref{fig:passive_no} also shows that the size distribution narrows
down for increasing height, because the larger ($a \ga 1\,\mu$m) particles
only make it to a height of about 10\,km, whereas the $a=0.1\,\mu$m
particles populate the atmosphere up to heights well above 50\,km with
nearly constant concentration, as already discussed in
Fig.\,(\ref{fig:n_n0}). The exact values for these results depend on
the diffusion constant assumed. If the diffusion constant was
10$\times$ larger than assumed in Eq.\,(\ref{eq:eddy_diff}), particles
with sizes up to $\sqrt{10}\approx 3\,\mu$m would populate heights up
to 10\,km.
%The absolute particle density $n_p\rm\,[cm^{-3}]$ falls
%with increasing height between 20\,km and 40\,km by about 1 order of
%magnitude, just like the gas density.  We find $n_p\approx\rm
%20\,cm^{-3}$ at 30\,km, a value slightly lower than measured by
%\cite{Knollenberg1980}.

Interestingly, the Grieger et al.\ data shows a re-increase of the
extinction coefficient above about 30\,km, whereas the
Knollenberg\,\&\,Hunten data show a roughly constant opacity at these
heights.  This could be an indication for an influx of particles from
above, likely from the sulphuric acid clouds, maybe seeded by the
impact of micro-meteorites at the top of the atmosphere
\citet{Gao2014}. If the particles are small and do not change size nor
composition, we expect an opacity slope that reflects the density
slope.  If there is a constant flux of particles settling down, we
expect a constant opacity, because in the Stokes regime $\vdreq$ is
independent of gas density (see Fig.\,\ref{fig:settle_growth} in the
Appendix).

\begin{figure*}
  % Series_03/Venus_active_full
  % .true.   ! viscous
  % Venus    ! gastype
  % DustChem_Venus2.dat  ! dustchem_file
  % 1        ! dust_diffuse
  % .true.   ! do_coagulation
  % 2        ! coagulation_Nbin
  % 0.3      ! precision
  % .false.  ! upwind_2nd_order
  % 101      ! Npoints
  % 1.5      ! zpot
  % 2        ! bc_high
  % .false.  ! detailed_charges
  % 50.0     ! q_over_a
  % .true.   ! power_correct
  \centering
  \vspace*{-3mm}
  \begin{tabular}{cc}
  \hspace*{-5mm}
  \includegraphics[page=10,height=63mm,width=80mm,trim=0 0 0 0,clip]
                  {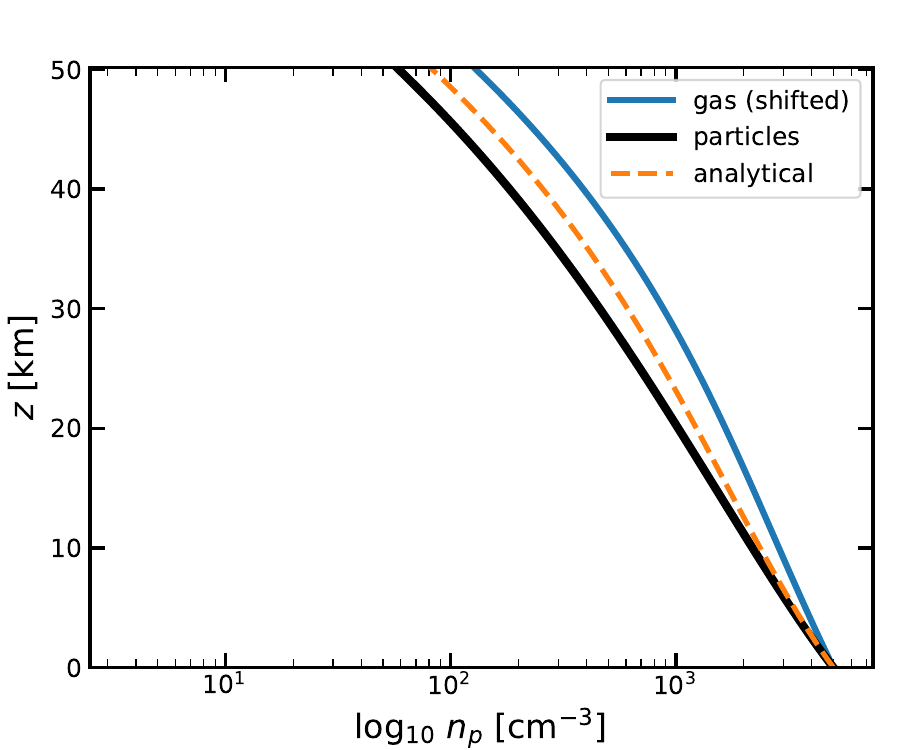} &
  \hspace*{-5mm}
  \includegraphics[page=9 ,height=63mm,width=80mm,trim=0 0 0 0,clip]
                  {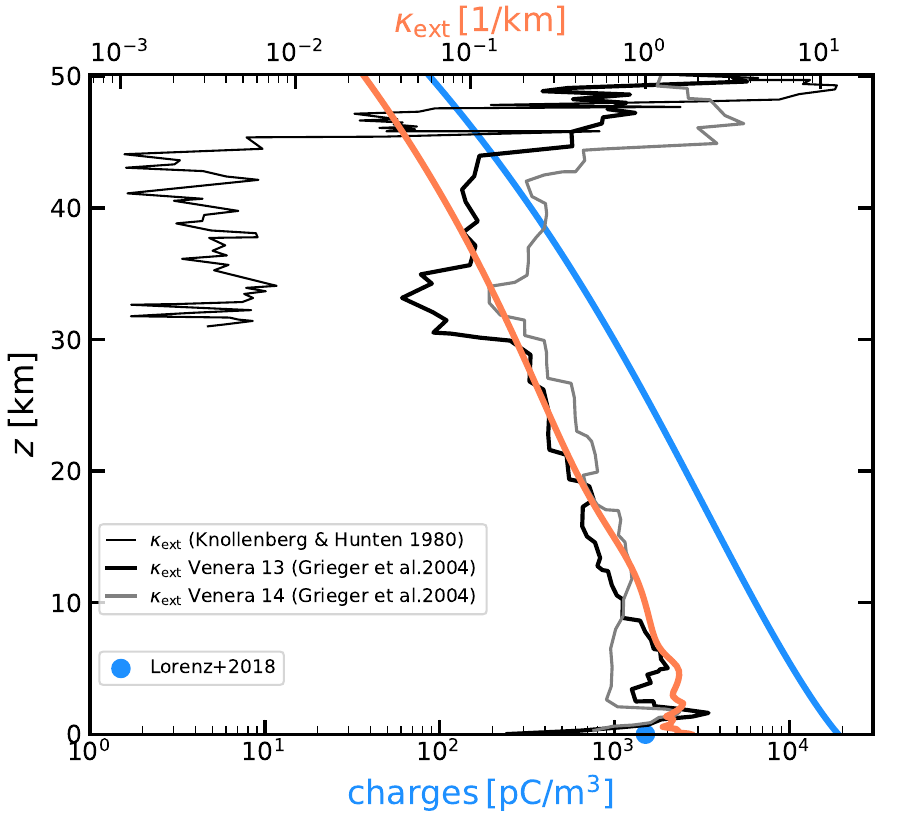} \\[-1mm]
  \hspace*{-5mm}
  \includegraphics[page=12,height=59mm,width=80mm,trim=0 0 0 28,clip]
                  {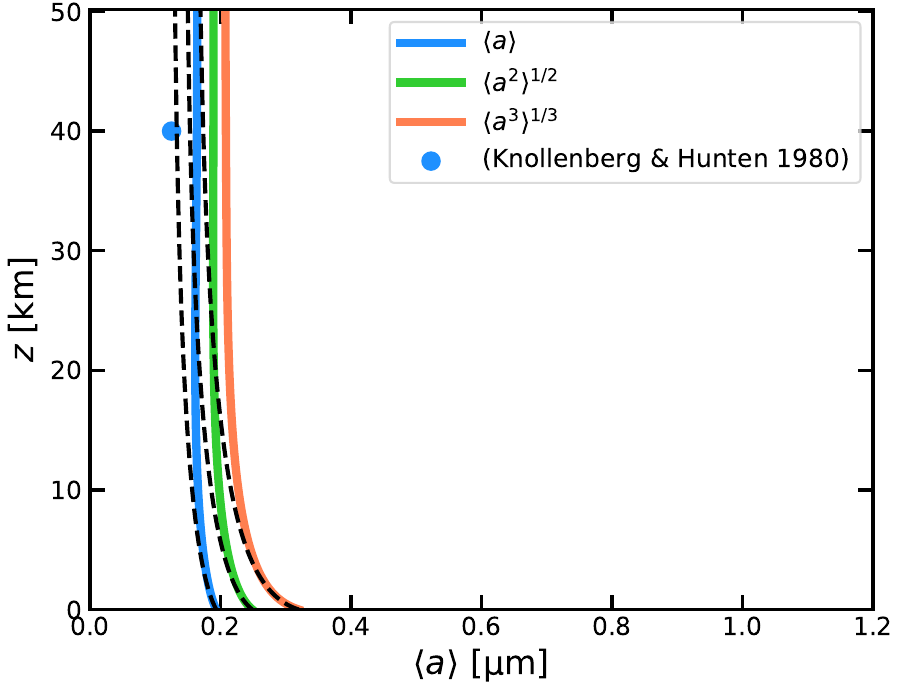} &
  \hspace*{-5mm}
  \includegraphics[page=15,height=59mm,width=80mm,trim=0 0 0 28,clip]
                  {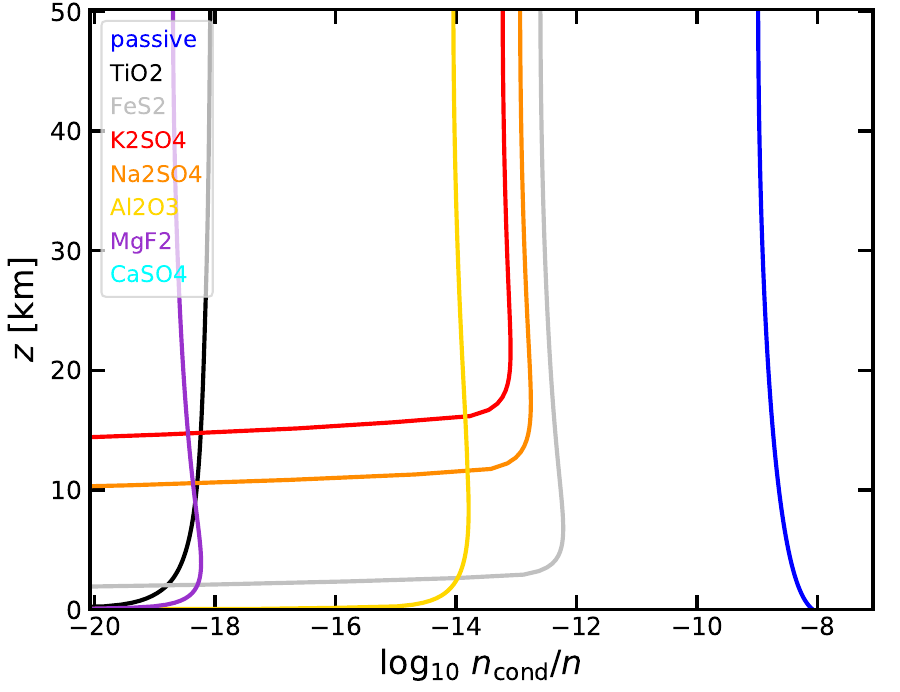} \\[-1mm]
  \hspace*{-5mm}
  \includegraphics[page=5 ,height=59mm,width=80mm,trim=0 0 0 28,clip]
                  {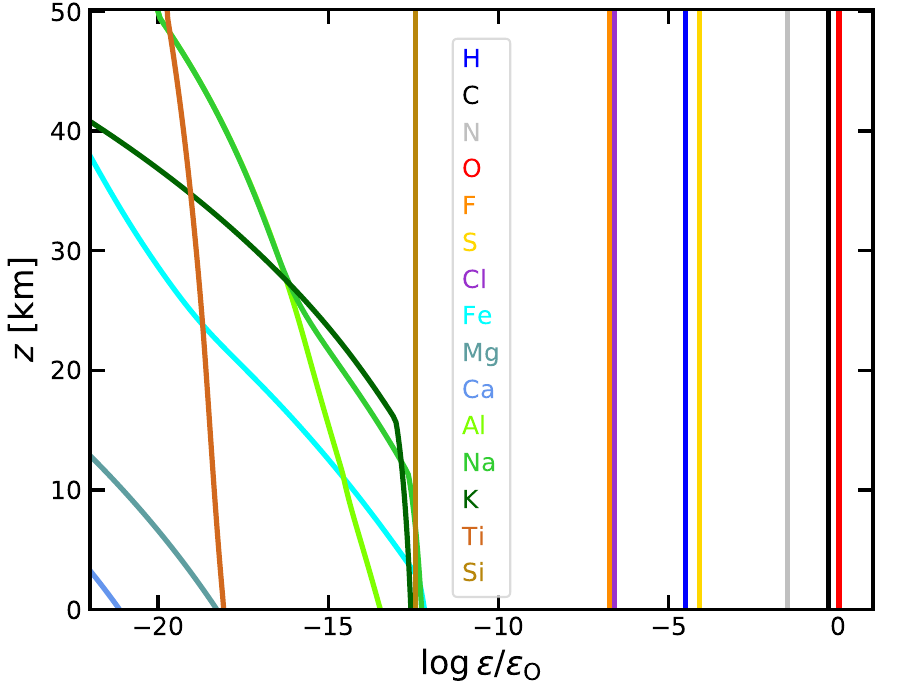} &
  \hspace*{-5mm}
  \includegraphics[page=7 ,height=59mm,width=80mm,trim=0 0 0 28,clip]
                  {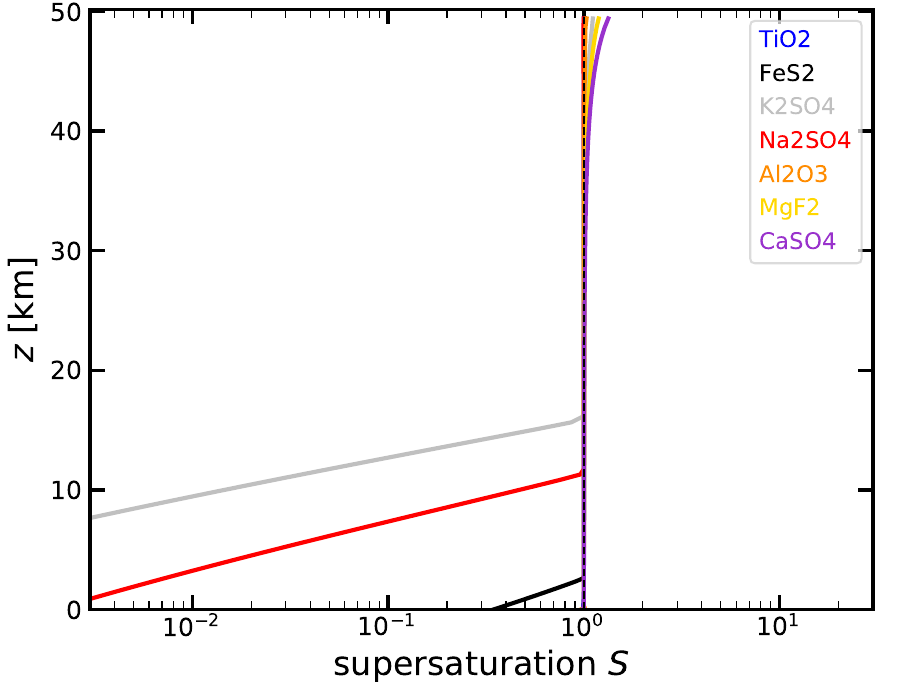} \\[-2mm]
  \end{tabular}
  \caption{Chemically active particles in the lower Venus atmosphere
    with coagulation, for grain charge parameter $\rm qa300=50$.}
    %$n_{\rm cond}$ is the number of condensed units per $\rm cm^3$.}
  \vspace*{1mm}
  \label{fig:active_full}
\end{figure*}

\begin{figure*}
  % Series_03/Venus_active_reduced
  % .true.   ! viscous
  % Venus    ! gastype
  % DustChem_Venus2.dat  ! dustchem_file
  % 1        ! dust_diffuse
  % .true.   ! do_coagulation
  % 0.3      ! precision
  % .false.  ! upwind_2nd_order
  % 101      ! Npoints
  % 1.5      ! zpot
  % 2        ! bc_high
  % .false.  ! detailed_charges
  % 5.0      ! q_over_a
  % .true.   ! power_correct
  % np=0.5 cm-3 instead of 5000 cm-3
  \vspace*{-3mm}
  \centering
  \begin{tabular}{cc}
  \hspace*{-5mm}
  \includegraphics[page=10,height=63mm,width=80mm,trim=0 0 0 0,clip]
                  {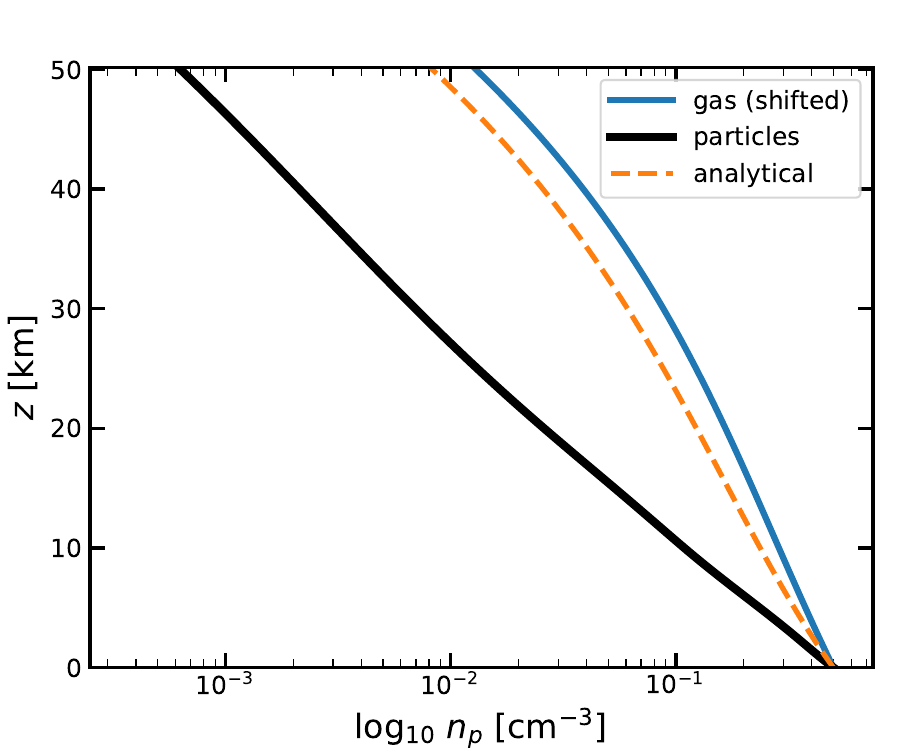} &
  \hspace*{-5mm}
  \includegraphics[page=9 ,height=63mm,width=80mm,trim=0 0 0 0,clip]
                  {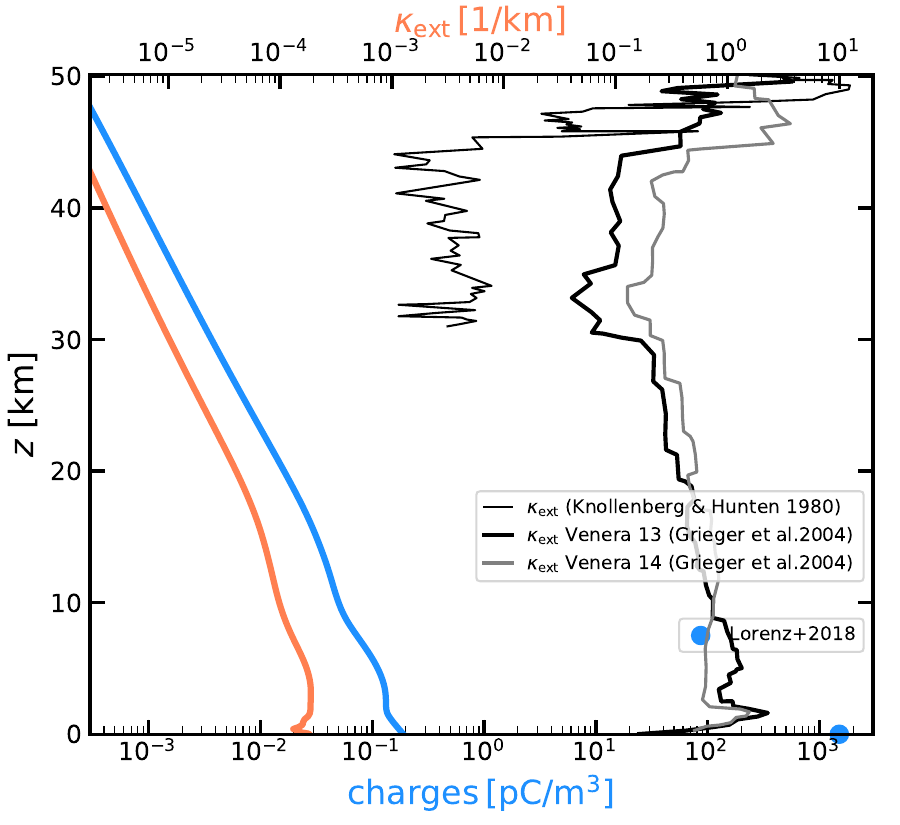} \\[-1mm]
  \hspace*{-5mm}
  \includegraphics[page=12,height=59mm,width=80mm,trim=0 0 0 28,clip]
                  {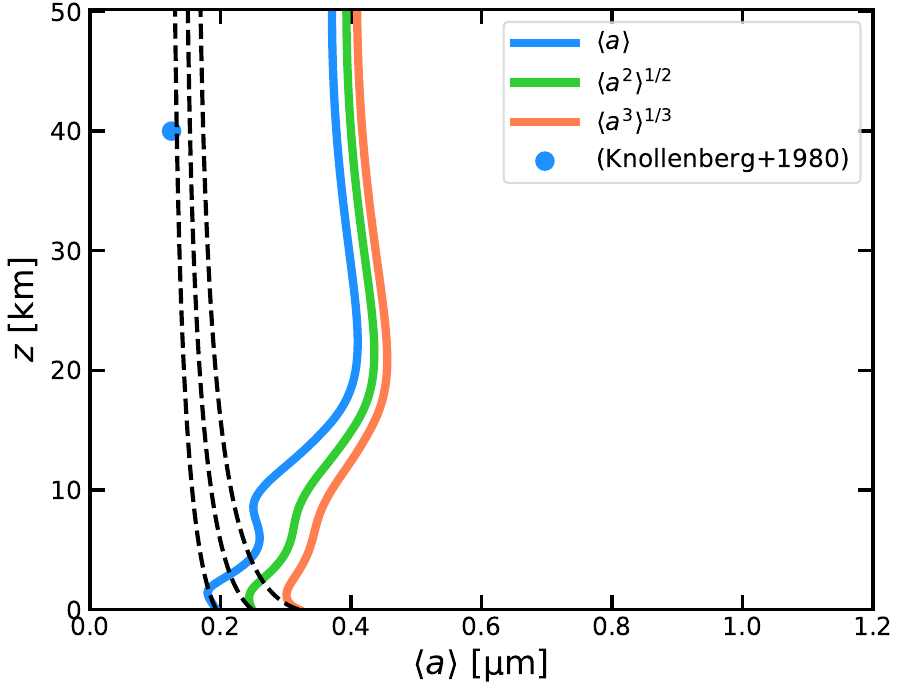} &
  \hspace*{-5mm}
  \includegraphics[page=15,height=59mm,width=80mm,trim=0 0 0 28,clip]
                  {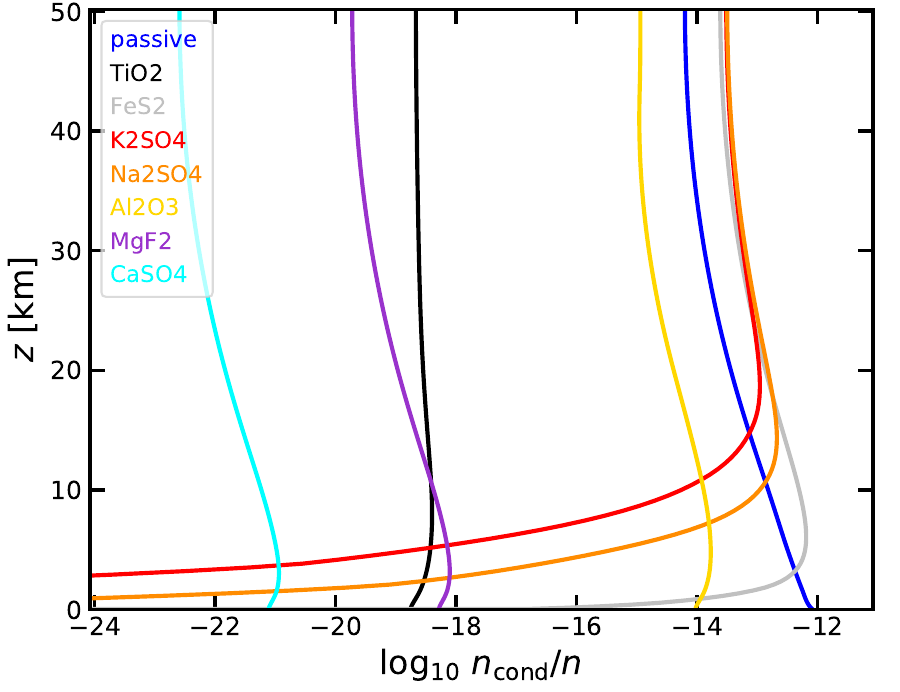} \\[-1mm]
  \hspace*{-5mm}
  \includegraphics[page=5 ,height=59mm,width=80mm,trim=0 0 0 28,clip]
                  {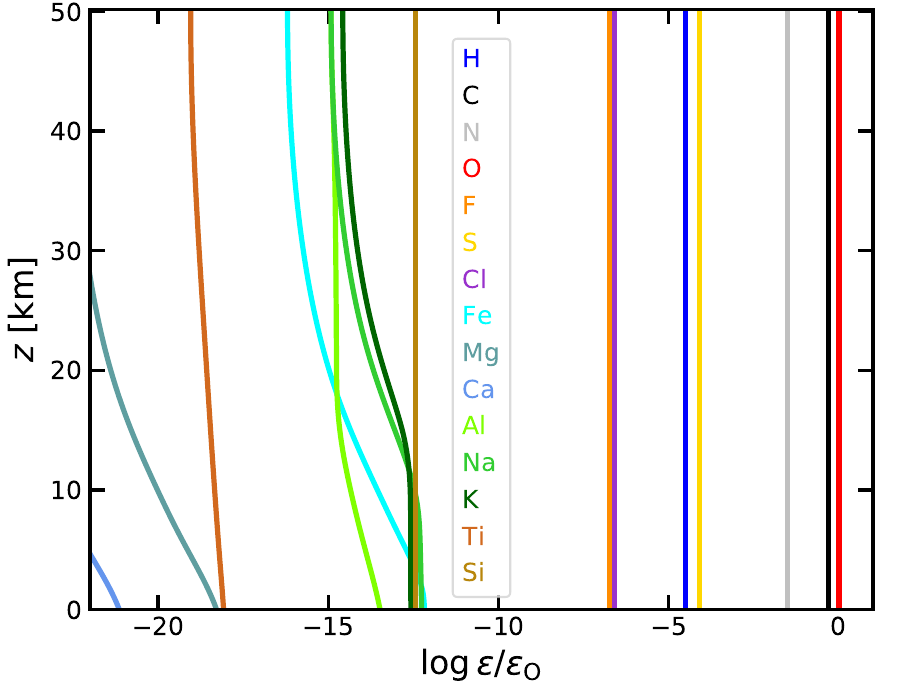} &
  \hspace*{-5mm}
  \includegraphics[page=7 ,height=59mm,width=80mm,trim=0 0 0 28,clip]
                  {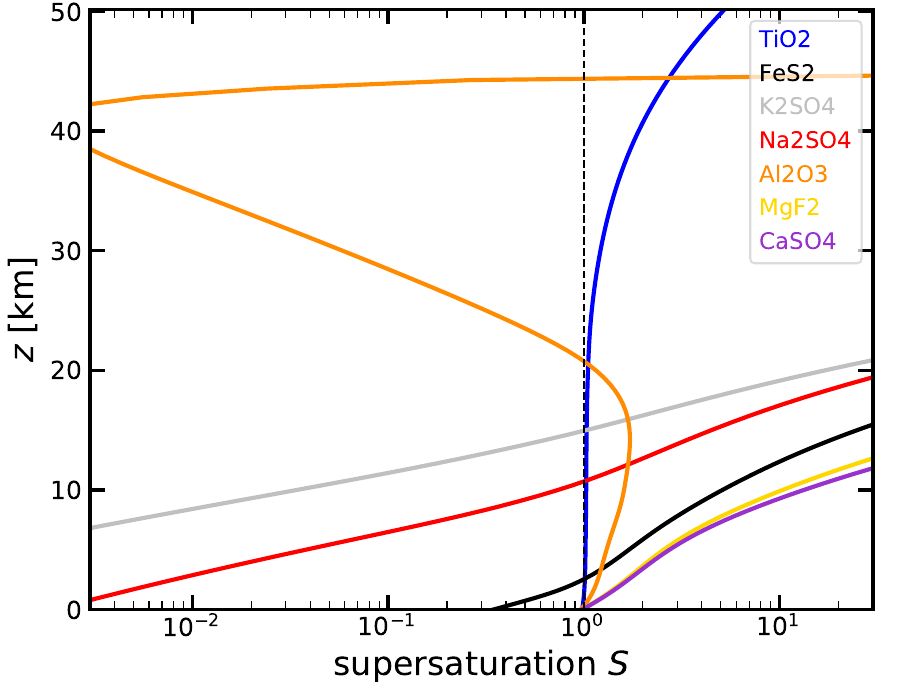} \\[-2mm]
  \end{tabular}
  \caption{Chemically active particles in the lower Venus atmosphere
    with coagulation and reduced passive material component, assuming
    only $n_p=0.5\rm\,cm^{-3}$ at the lower boundary. The grain charge
    parameter is set to $\rm qa300=5$.}
  \vspace*{1mm}
  \label{fig:active_reduced}
\end{figure*}

Figure \ref{fig:passive_yes} shows the same model with coagulation
switched on, in comparison to the same analytic model without
coagulation as shown in the previous figure, for various grain charge
parameter values.  In this model, the particles collide to form fewer
and bigger particles that settle more quickly, leading to a steep
gradient of $n_p(z)$ close to the bottom of the atmosphere if the
particles are weakly charged.  However, for grain charge parameter
$\ga 30$, the coagulation becomes increasingly inefficient, because
the electrostatic repulsion between the particles prevents their
collisions. Eventually, for large grain charge parameters $\ga 50$,
the results resemble the model without coagulation as shown in
Fig.\,\ref{fig:passive_no}.  For medium charges, the particles
form fewer and bigger ($a>0.3\,\mu$m) particles, which is in
conflict with both the measured opacities and mean sizes of the
particles at 40\,km height.  Thus, our model suggests that the
particles in the lower Venus atmosphere are strongly charged ($\rm
qa300\!>\!50$), which is actually in agreement with our physical
expectations, see App.\,\ref{AppC}.

However, the total charge density of the particles becomes too high when
we increase qa300 from 4 (where it fits the Lorenz et al.\ data)
to 50.  We conclude that it is currently not possible with our model
to fit both the measured opacity and the charge density data.

%The slightly larger particles do populate the entire lower Venus
%atmosphere, but in comparison to the model without coagulation, we
%find about 5$\times$ fewer particles.  Coagulation is particularly
%important in the lower part of the lower Venus atmosphere, because the
%coagulation rates scale with $\rho^2$.

\subsection{Active particles}
\noindent
Figures \ref{fig:active_full} and \ref{fig:active_reduced} show our
results for chemically active particles including coagulation. The
models have been set up with the 7 materials and surface reactions
listed in Table~\ref{tab:conc}, one additional chemically unspecified
``passive'' material component, and 15 elements (H, C, N, O, S, F, Cl,
Fe, Mg, Ca, Al, Na, K, Ti, Si), out of which 10 are involved in the
condensation process. The GGchem-part of {\sc DiffuDrift v2} finds 308
molecules in its database for this choice of elements. Besides the
particle quantities already introduced and plotted in our models for
passive particles, we now show in addition the gas element abundances
$\epsilon_k$ as function of height, the amount of condensed solid
units per gas particle $n_{\rm cond}/n$ and the supersaturation ratios
$S$ of the various materials.

For the model shown in Fig.~\ref{fig:active_full} we have chosen the
same lower BC with $n_p(z\!=\!0)=5000\rm\,cm^{-3}$ as in the previous
models for passive particles, and grain charge parameter $\rm
qa300=50$. The particle densities and sizes found in this model are
very similar to the results of the same model for passive
particles. Close inspection shows, however, that the particles have
grown to very slightly bigger sizes due to the deposition of pyrite
\ce{FeS2[s]}, sodium sulphate \ce{Na2SO4[s]} and potassium sulphate
\ce{K2SO4[s]} on their surfaces.  These coatings cause the vertical
absorption optical depth to increase very slightly, from 0.49 to 0.52.
However, when plotting their concentration $n_{\rm cond}/n$ as
function of height, the deposits do not show a saw-tooth-like profile
as predicted by our simple GGchem equilibrium condensation model
(Fig.\,\ref{fig:ggchem}), but extend all the way up to the upper
boundary of the model with about constant concentrations. This is
because once these depositions have formed, they do not come off again
easily, as the upward transport by diffusion is faster than the
sublimation.

However, these depositions only amount to a layer of an average $\rm
thickness \approx 0.3$\AA, i.e.\ not even a single mono-layer. This is
a straightforward consequence of the availability of condensible
elements (or order $10^{-12}$) compared to the passive particles as
set by our lower BC, which translates to a chemical abundance of about
$10^{-8}$.  Therefore, the results of this model are physically not
very meaningful.  However, it is still reassuring to see that (a) the
condensations take place at the heights predicted by our simple GGchem
equilibrium condensation model (Sect.\,\ref{sec:GGchem}), (b) all
supersaturation ratios are limited by about 1, and (c) the gas element
abundances of the condensing metals are reduced in the upper regions
of the model as expected from our GGchem model, i.e. the condensible
molecules have sufficient time to find a surface to deposit.

In order to study the behaviour of chemically active particles in more
detail, we have computed another model (shown in
Fig.~\ref{fig:active_reduced}) where the abundance of the passive
particles at the bottom of the atmosphere is reduced by 4 orders of
magnitude, using $n_p(z\!=\!0)=0.5\rm\,cm^{-3}$ instead of
$5000\rm\,cm^{-3}$, such that the aforementioned abundances of
active and passive materials become about equal.  In this model,
coagulation is much less important, because it scales with $n_p^2$.
To bring all four principle timescales (settling, diffusion, growth
and coagulation) to about the same order of magnitude, we have chosen
to reduce the grain charge parameter to $\rm qa300=5$.

This model (Fig.~\ref{fig:active_reduced}) shows a rather different
dynamical and chemical behaviour of the active particles. The
depositions of \ce{FeS2[s]}, \ce{Na2SO4[s]} and \ce{K2SO4[s]} cover
all particles with a layer of $\rm thickness\approx 0.2\,\mu$m,
i.e. about 1000 mono-layers. These larger particles settle
efficiently, which reduces $n_p(z)$ considerably in the upper
layers. The material mixture in the upper layers is characterised by
about equal amounts of these three condensates, followed by the
passive component and by \ce{Al2O3[s]}, with some traces of
\ce{TiO2[s]}, \ce{MgF2[s]} and \ce{CaSO4[s]}.  This behaviour is due
to the incompleteness of the condensation process in this model: (i)
The thermal velocities of the key molecules \ce{FeCl2}, \ce{NaCl} and
\ce{KCl} are about equal, and therefore, about equal amounts of
\ce{FeS2[s]}, \ce{Na2SO4[s]} and \ce{K2SO4[s]} are found to condense,
(ii) the abundances of the condensible elements in the gas phase does
not fully follow the pattern of the GGchem equilibrium condensation
model, and (iii) the gas remains highly supersaturated in the upper
layers.

Close inspection of $n_{\rm cond}/n$ shows that the downward transport
of the particles by diffusion and settling is comparable to the speed
of sublimation, and hence the depositions on these particles, while
shrinking, still exist below their nominal cloud bases, where they are
undersaturated according to our simple GGchem equilibrium condensation
model.  We note that the particles are always mixed by diffusion in
this model, so there are no sharp edges as in the equilibrium
condensation {\sc GGchem} model.

\begin{figure*}
  % Series_03/Venus_passive 
  % Series_03/Venus_passive_coag (q/a=70)
  % Series_03/Venus_active_full (q/a=100)
  % Series_03/Venus_active_reduced (q/a=10)
  \centering
  {\sf passive particles}\\
  \includegraphics[height=47mm,width=135mm,trim=15 52 0 7,clip]
                  {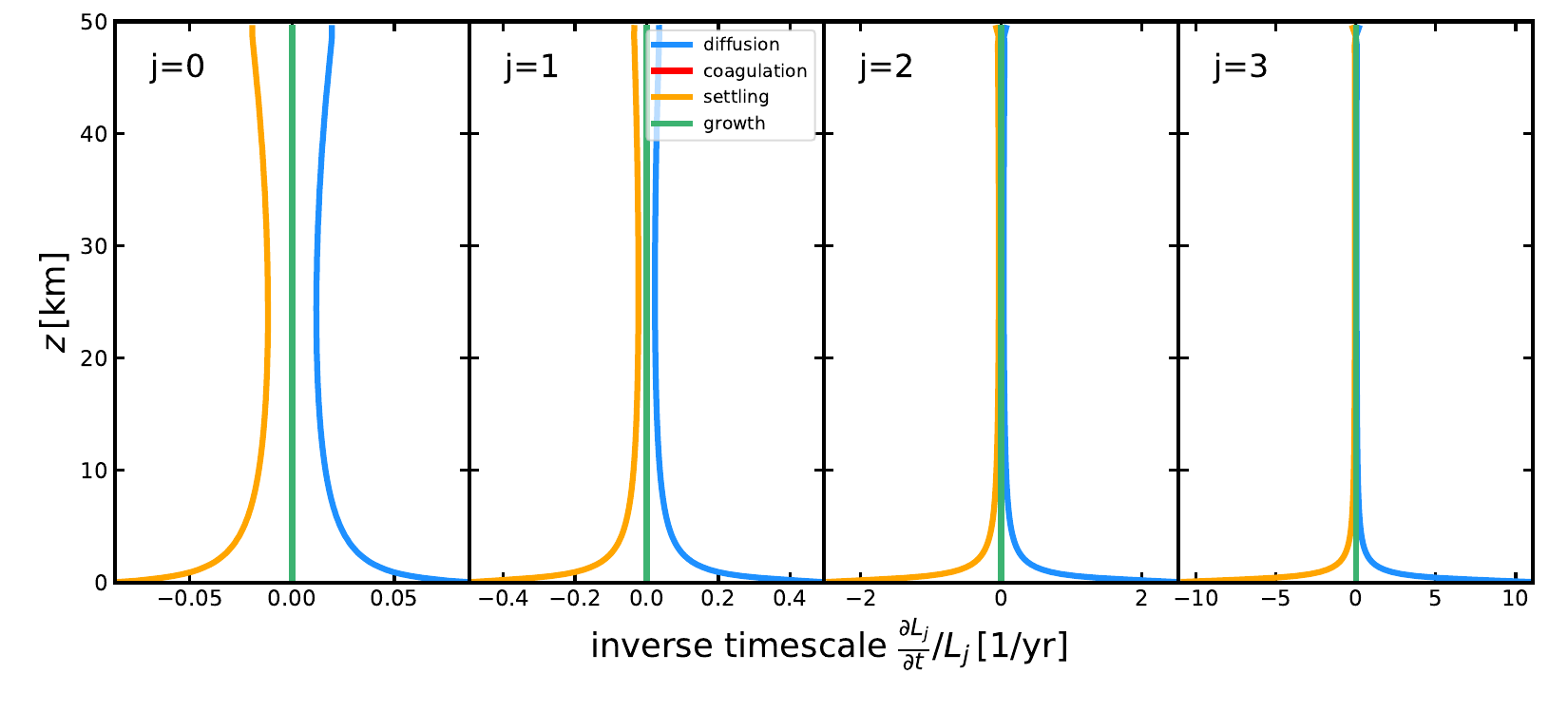}\\[-1mm]
  {\sf passive particles with coagulation}\\
  \includegraphics[height=47mm,width=135mm,trim=15 52 0 7,clip]
                  {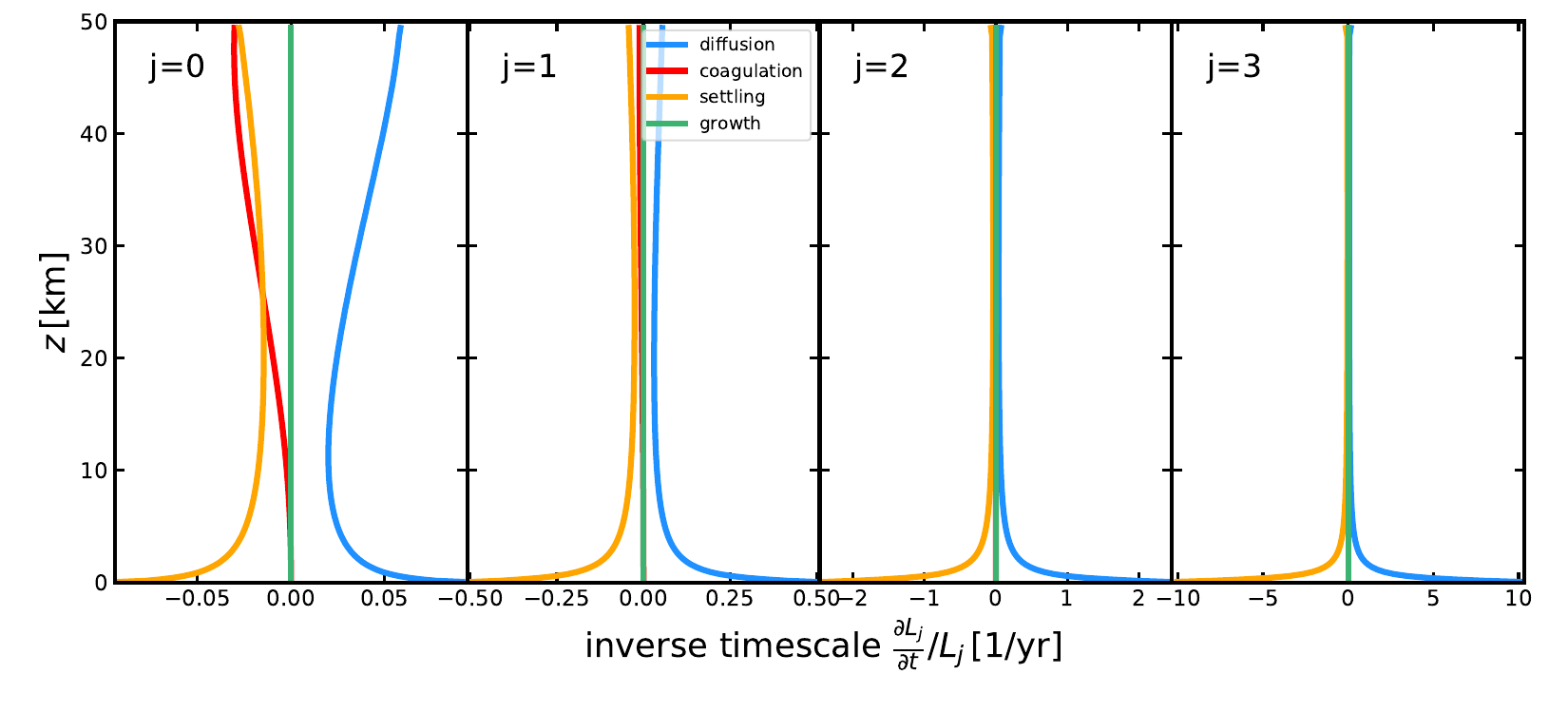}\\[-1mm]
  {\sf active particles with coagulation}\\
  \includegraphics[height=47mm,width=135mm,trim=15 52 0 7,clip]
                  {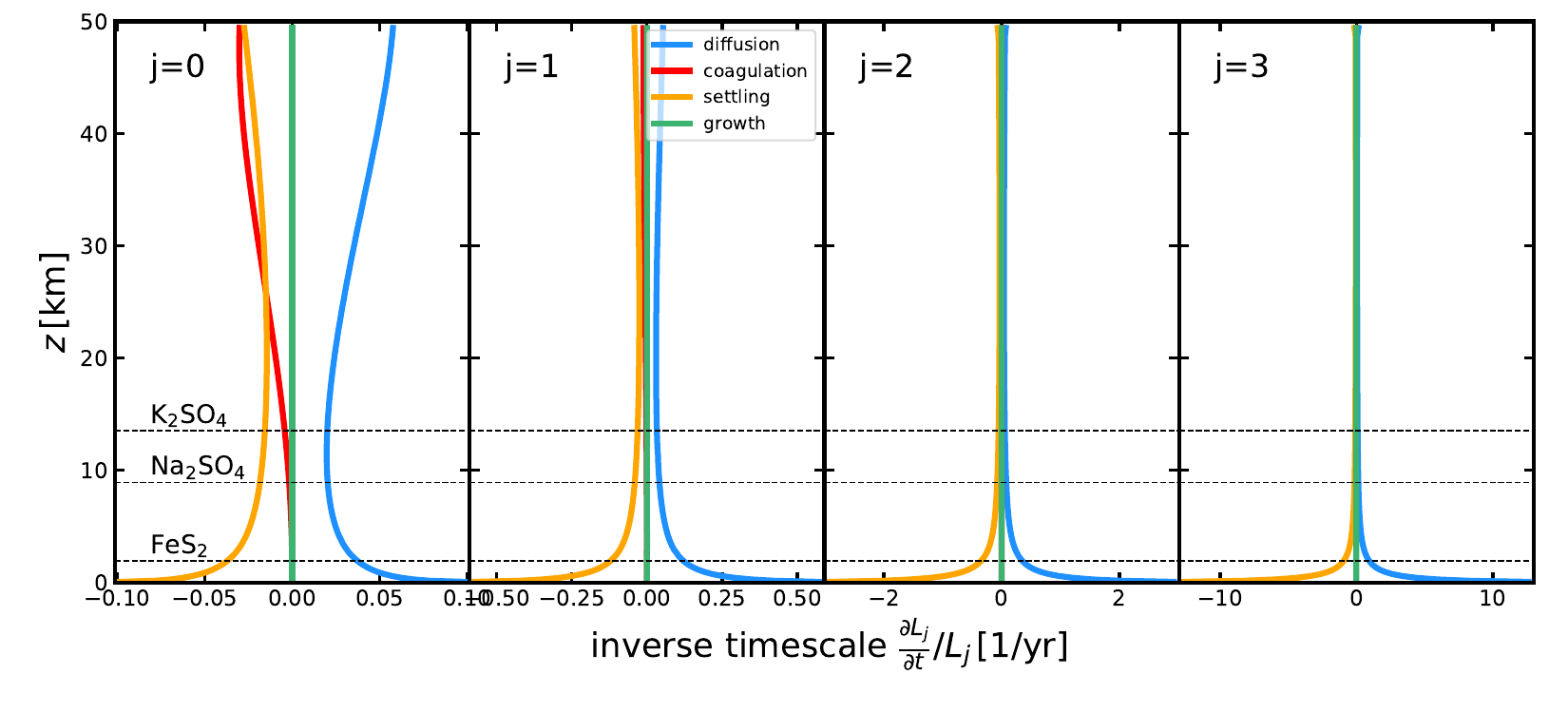}\\[-1mm]
  {\sf active particles with coagulation \& reduced passive component}\\
  \includegraphics[height=52mm,width=135mm,trim=15 20 0 7,clip]
                  {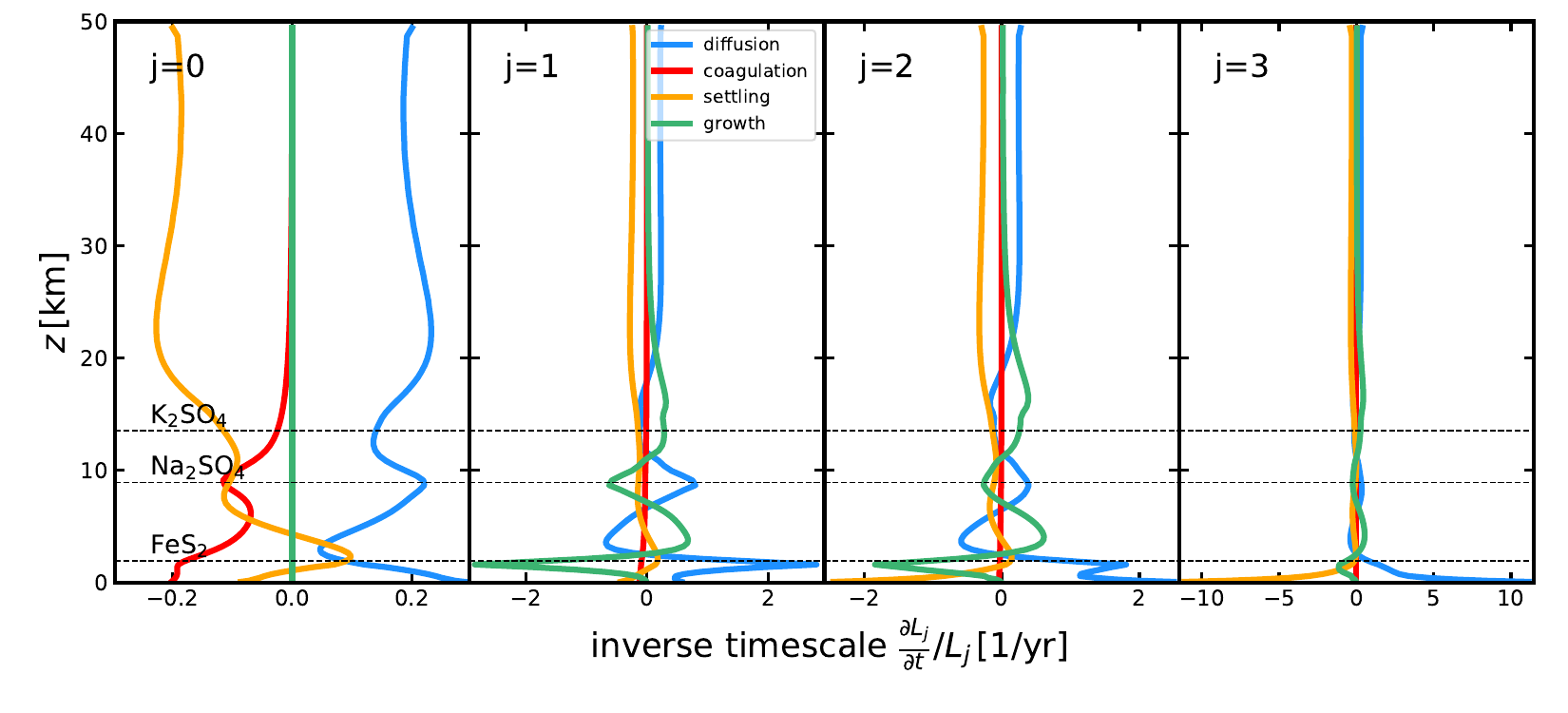}\\[-1mm]
  \caption{Timescale analysis of our four models as indicated. The
    top model is for passive particles without coagulation, see
    Fig.\,\ref{fig:passive_no}.  From the models for passive particles
    with coagulation (Fig.\,\ref{fig:passive_yes}), the model with
    $q/a=50$ is selected here.  The two lower models for active particles
    correspond to Figs.\,\ref{fig:active_full} and \ref{fig:active_reduced},
    respectively. The graphs show the inverse characteristic
    timescales $\rm[1/yr]$ of each cloud particle moments
    $L_j\,(j\!=\!0,1,2,3)$ for the four basic physical processes
    included: diffusion (blue), coagulation (red), settling (orange)
    and net growth (green). At every height, the sum of
    these inverse timescales must be zero in steady state.  In the two
    lower models for chemically active particles, we have added
    horizontal lines at 1.9\,km, 8.9\,km and 13.5\,km to mark the
    cloud bases of pyrite, Na-sulphate and K-sulphate, respectively.}
  \label{fig:timescales}
  \vspace*{-5mm}
\end{figure*}

\subsection{Timescale analysis}
\noindent
Figure~\ref{fig:timescales} shows the relevance of the four basic
processes diffusion, coagulation, settling and net growth by means of
a timescale analysis.  We have calculated inverse timescales as
$(\Delta L_j/\Delta t)/L_j\rm\,[yr^{-1}]$ from the results of the
respective operators in {\sc DiffuDrift} at the final timestep $\Delta
t$.  At any height in the atmosphere, the sum of these inverse
timescales must be zero in steady state.

In most cases, there are two leading processes that cancel each other,
for example in the most basic model for passive particles without
coagulation at the top of Fig.~\ref{fig:timescales}, diffusion
replenishes the particles within about two weeks, and settling removes
them at the same rate.  In the model for passive particles with
coagulation (second model from the top), we see that coagulation
becomes increasingly more relevant in removing the particles at higher
altitudes where the particles are less charged, and the gain by
diffusion is balanced by a combination of settling and coagulation, in
consideration of the number of the particles ($\to L_0$).  However,
since coagulation has no effect on the total condensed volume of the
particles ($\to L_3$), the diffusion is still balanced by settling
when considering $j\!=\!3$.

The timescale analysis of the third model for active particles is
practically identical to the second model for passive particles,
because the amount of the additional depositions by the active
condensates is negligible compared to the passive material component.
However, the bottom model for chemically active particles with reduced
passive component reveals new insights.  Growth\,\&\,evaporation have
no effect on the number of particles $(L_0)$. However, for all
$j\!>\!0$, the net growth becomes relevant, turning negative where one
material on the particles sublimates.  We mark the negative green
peaks of the sublimation rates of \ce{K2SO4}, \ce{Na2SO4} and
\ce{FeS2} with additional height lines at 13.5\,km, 8.9\,km and
1.9\,km, respectively.

Between these negative sublimation peaks, the inverse growth timescale
is positive as gas molecules continue to deposit on the particles'
surfaces.  The wiggles in the growth timescale are counterbalanced by
corresponding anti-wiggles in the diffusion timescale, i.e.\ the
liberated molecules are transported away by diffusion, whereas the
settling and coagulation timescales remain smooth.

\section{Discussion}
\subsection{Element trapping in clouds}
\noindent
This paper has made some predictions about the material composition of
aerosol particles in the lower Venus atmosphere based on two model
approaches of increasing complexity, an equilibrium condensation model
(Sect.~\ref{sec:GGchem}) and a dynamical particle model with kinetic
condensation (Sect.~\ref{sec:DiffuDrift}).  Both models agree, that
elements can be trapped in cloud layers, such that the abundance of an
element above a cloud layer is significantly reduced with respect to
its abundance below the clouds, provided that this element is consumed
by the formation of the cloud particles.  In this section, we want to
shed some light on the microphysics behind this important process in
planetary atmospheres.

We multiply Eq.\,(\ref{eq:L3s}) by $\nu_{s,k}/V_0^s$ and sum up these
equations for all materials $s$. We then add Eq.\,(\ref{eq:epsk}).
This way, the chemical source terms (nucleation and net growth) cancel
out, and we find in the stationary limit
\begin{align}      
  &\frac{\partial}{\partial z} \Bigg(\sum_s \frac{\nu_{s,k}}{V_0^s}
  \sum_{i=1}^2 n_i V_i^s\,\vdreq_{,i}\Bigg) \nonumber\\
  &+\,\frac{\partial}{\partial z}\,
  \Bigg(D\,\rho\,\frac{\partial}{\partial z}
  \bigg(\sum_s\frac{L_3^s\,\nu_{s,k}}{V_0^s}+\epsilon_k^{\rm gas}\bigg)\Bigg)
  ~=~ 0 \ .
\end{align}
Identifying $\,\epsilon_k^{\rm cond}=\sum_s\frac{\nu_{s,k}}{V_0^s}\,L_3^s
= \frac{1}{\rho}\sum_s \frac{\nu_{s,k}}{V_0^s} \sum_{i=1}^2 n_i V_i^s\,$ as
the abundance of element $k$ in condensed form, i.e.\ the number of
nuclei of element $k$ in the condensed particles per gram of gas. Using
Eq.\,(\ref{eq:meanvd}), we find the following remarkable equation
\begin{equation}
  \rho\,\epsilon_k^{\rm cond} \langle \vdreq^{\!\!3}\rangle
  +\,D\,\rho\,\frac{\partial}{\partial z}
  \Big(\epsilon_k^{\rm cond}+\epsilon_k^{\rm gas}\Big)
  ~=~ \mbox{const} \ ,
  \label{eq:epsflux}
\end{equation}
where $\langle\vdreq^{\!\!3}\rangle$ is the volume-mean settling
velocity, see Eq.\,(\ref{eq:meanvd}).
The constant in Eq.\,(\ref{eq:epsflux}) is the time and
height-independent flux of element $k$ through the atmosphere
$\rm[nuclei/s/cm^2]$.  Assuming this constant flux to be zero (e.g.\ no
micro-meteoritic influx) we find the following three limiting cases
\begin{align}
  &&&&\nonumber\\*[-9mm]
  \mbox{case 1} &&
  \big(\epsilon_k^{\rm cond}\gg\epsilon_k^{\rm gas}\big)&:&
  \displaystyle
  \frac{\partial}{\partial z} \ln \epsilon_k^{\rm cond}
  = -\,\frac{\langle\vdreq^{\!\!3}\rangle}{D}
  \label{eq:case1}\\ 
  \mbox{case 2} &&
  \big(\epsilon_k^{\rm gas}\gg\epsilon_k^{\rm cond}\big)&:&
  \displaystyle\hspace*{-2mm}
  \frac{\partial}{\partial z} \epsilon_k^{\rm gas}
  = -\,\frac{\epsilon_k^{\rm cond} \langle\vdreq^{\!\!3}\rangle}{D}
  \label{eq:case2}\\
  \mbox{case 3} && \hspace*{-2mm}
  \big(D\gg\langle\vdreq^{\!\!3}\rangle H_p\big)&:&
  \displaystyle
  \epsilon_k^{\rm cond}+\epsilon_k^{\rm gas} = {\rm const}
  \label{eq:case3} \ .\\*[-5mm]
  &&&&\nonumber
\end{align}
The first case (Eq.\,\ref{eq:case1}) corresponds to the behaviour of
passive particles, as already derived in Sect.\,\ref{sec:passive}
(Eq.\,\ref{eq:passive}), stating that the concentration of particulate
matter can only decrease with height, which it does in particular when
the mixing is insufficient.  The second case (Eq.\,\ref{eq:case2}) is
discussed in detail by \cite{Woitke2020}. It states that gas element
abundances decrease with height in a cloud layer, see r.h.s.\ of
Fig.\,\ref{fig:ggchem} and lower left plots in
Figs.\,~\ref{fig:active_full} and \ref{fig:active_reduced}. When
elements rain out, a negative $\epsilon_k^{\rm gas}(z)$-gradient must
build up in the gas phase to cause an upward-directed diffusive
element flux that counteracts the settling flux.
Equation~(\ref{eq:case2}) states explicitly that gaseous element
abundances can never increase with height in static planetary
atmospheres.  The third case (Eq.\,\ref{eq:case3}) shows, however,
that this is actually possible in very well-mixed cases, for example
high in the atmosphere, when a $T$-inversion causes clouds to turn
back into gases. For this case to occur, we need
$V_g\!\gg\!\langle\vdreq^{\!\!3}\rangle$, where $V_g\!=\!D/H_p$ is the
vertical gas mixing velocity, see Eq.\,(\ref{eq:Vg}).

The third case does not occur in this model, but the atmospheric
pattern evolves from case~2 (Eq.\,\ref{eq:case2}) directly into case~1
(Eq.\,\ref{eq:case1}), as soon as the reservoir of metals like Fe, Na
and K is exhausted with increasing height. Since the particles stay
very small ($a\!<\!0.2\,\mu$m above 20\,km), they remain coupled to
the gas and the material composition of the particles does not change
much upward of 20\,km.  This behaviour is only revealed by the kinetic
condensation models, whereas the material composition of the particles
keeps changing with height in the equilibrium condensation models.

%-------------------------------------------------------------------
\section{Summary and discussion}
%-------------------------------------------------------------------
\label{sec:summary}
\noindent
This paper has made a number of predictions concerning the gas phase
chemistry and the properties and material composition of
sub-$\mu$m-sized aerosol particles in the lower part of the Venus
atmosphere, below the main sulphuric acid cloud layer at a height of
about 45\,km.

The {\sc GGchem} model discussed in Sect.~\ref{sec:GGchem} uses the
same setup as published in \cite{Rimmer2021} for the interface between
the surface and the gas at the bottom of the atmosphere. Assuming
chemical equilibrium and phase equilibrium, a number of metal-chloride
and metal-fluoride molecules are found to be present in the gas phase
over the Venus surface, in particular \ce{FeCl2}, \ce{NaCl}, \ce{KCl},
\ce{SiF4}, \ce{AlF2O}, \ce{TiF4}, \ce{MgCl2}, and \ce{CaCl2}, with
very low concentrations of order $10^{-12}$ to $10^{-21}$.  The
formation of these molecules is a consequence of chlorine and fluorine
being available in the Venus atmosphere in form of HCl and HF
molecules.

The trace concentrations of these metal molecules are sufficient to
stabilise a number of solid materials, which can deposit on aerosol
particles in the lower Venus atmosphere.  In particular we predict
potassium sulphate \ce{K2SO4[s]} to form above a height of about
15.5\,km, and sodium sulphate \ce{Na2SO4[s]} above 9.5\,km.  We call
these condensations sulphate hazes.  In addition, pyrite \ce{FeS2[s]}
can deposit above 2.4\,km as already noted by \cite{Byrne2024}. These results are in
close agreement with the predictions from our {\sc GGchem} model
(13.8\,km, 9.3\,km, and 2.9\,km, respectively), and coincide well with
the three potential dust layers found in the Pioneer Venus Large Probe
neutral mass spectrometer data by \cite{Mogul2023}. 
\ce{Fe2O3[s]} {\sl (hematite)},
\ce{CaSO4[s]} {\sl (anhydrite)},
\ce{Al2O3[s]} {\sl (corundum)},
\ce{TiO2[s]} {\sl (rutile)}, and
\ce{MgF2[s]} {\sl (magnesium ﬂuoride)} are found to be
stable in the gas as well, all the way down to the surface.
{We note that the chemical pre-conditions for condensations in the lower Venus atmosphere are distinctively different from the conditions in the Venus surface, because certain elements like Si are entirely confined in the surface and no longer available to form minerals in the atmosphere.}

To model the behaviour of the aerosol particles, we have improved the
{\sc DiffuDrift} code originally developed by \cite{Woitke2020} for
brown dwarf and exoplanet atmospheres. We have re-formulated the basic
equations to allow for a proper treatment of settling and growth at
arbitrary Knudsen numbers, including the limiting cases in the Epstein
and Stokes regimes.  We have included coagulation driven by Brownian
motion, difference in settling velocities and turbulence, taking into
account the repelling effect of particle charges. We have furthermore
added an unspecified, chemical inert passive material that forms the
core of the aerosol particles, and have improved the numerics.

We provide analytic solutions for the most simple case of passive
particles with a given concentration at the bottom of the atmosphere,
only affected by diffusive mixing and settling
(Eq.\,\ref{eq:passive}).  Particles with radius $a>1\,\mu$m cannot get
to heights $>10\,$km, but sub-micron particles ($a \la
0.3\,\mu$m) stay well-mixed with the gas and can reach the main
sulphuric acid clouds from below.  The exact values depend on the eddy
diffusion coefficient assumed.  For example, if the diffusion
coefficient was $10\times$ larger than assumed, then $\sqrt{10}\times$
larger particles would reach the same height.

According to our models for chemically active particles, we expect the
aerosol particles in the lower Venus atmosphere to be covered by a
thin layer of \ce{K2SO4[s]}, \ce{Na2SO4[s]} and \ce{FeS2[s]}. At a
height of 45\,km, we reach particle radii of about 0.15 to
0.25\,$\mu$m and particle densities of about 10 to $\rm 100\,cm^{-3}$,
depending on the assumptions about the near-surface aerosol particle
concentration and size distribution, and the efficiency of coagulation
connected to the particle charges.  We conclude that the particles
must be strongly charged negatively, or order 100 negative charges per
micron of particle radius, otherwise our model would produce steep
gradients above the surface, which is inconsistent with the
\cite{Grieger2004} opacity data and the \cite{Lorenz2018} discharge
current data.

In a model like ours, where particles are dredged up from the ground,
and there is an equilibrium between upward mixing and gravitational
settling, the particle concentration must decrease with height at
least as fast as the gas density, which is in good agreement with the
slope of the \cite{Grieger2004} opacity data below 35\,km.  However,
above that height, the data shows an increasing trend, which suggests
that particles are inserted from above and travel down the atmosphere,
meeting the dredged-up particles at around 35\,km height. Since the
fall velocity is independent of gas density in the Stokes regime, a
constant influx of particles translates into a constant particle
density, i.e.\ an increasing particle concentration with increasing
height. {Such a particle influx could be provided by the impact
of micro-meteorites as discussed in \cite{Gao2014}. Another
possibility is a horizontal transport of cloud particles, which then
shrink to their refractory core at the sulphuric acid cloud base and
continue to settle down slowly as aerosol particles.  These effects
will be investigated in a forthcoming paper.}

\begin{acknowledgements}
      We thank the anomynous referee for his detailed list of
      additional references and suggestions to improve the paper,
      in particular with regard to questions about the surface mineral
      composition and the chemistry in the near-surface atmosphere of
      Venus.  PW, MS, HL and FW thank the Austrian Science Fund (FWF)
      for the support of the VeReDo research project, grant I6857-N.
      MF, KN, PE and JK thank the Czech Science Foundation (GACR) for
      the support of the VeReDo research project, grant I6857-N.  TC
      thanks the Science and Technology Facilities Council (STFC) for
      the PhD studentship (grant reference ST/X508299/1).
\end{acknowledgements}

\bibliographystyle{aasjournalv7}
\bibliography{references}

%======================================================================

\begin{appendix}

\section{Gravitational settling}
\label{AppA}
\noindent
Using the double-$\delta$ representation of $f(V)$ (Eq.\,\ref{eq:double-d}), we can
express the divergence of the vertical flux of cloud particles
settling gravitationally in the atmosphere as
\begin{equation}
  \frac{\partial(\rho L_j)}{\partial t}\bigg|_{\rm settle}\!=~
  \frac{\partial}{\partial z}\sum_{i=1}^2 n_i\,V_i^{\;j/3}\,\vdreq_{,i}
  %~=~ \frac{d}{dz}\left(\langle \vdreq^{\!j}\rangle \rho L_j\right)
  \label{eq:settle}
\end{equation}
where $\vdreq_{,i}$ is the downward equilibrium drift velocity (or
final fall speed) of the representative particle $i$, which generally
depends on particle radius $a$ and gas density $\rho$ in complicated
ways, see Sect.~2.3 in \cite{Woitke2003}. The main results
of this paper are repeated here in the following.

\subsection{Large Knudsen numbers (Epstein regime)}
\noindent
In the Epstein regime, the equilibrium drift velocity is given by
\cite{Schaaf1963} as
\noindent
\begin{equation}
  \vdreq = g\,\frac{\rho_{\rm m}\,a}{\rho\,\vth}
\end{equation}
where $g$ is the gravitational acceleration, $\rho_{\rm m}$ the
material density of the particles,
$\vth\!=\!\big(\frac{8\,kT}{\pi\,\mu}\big)^{1/2}$ the thermal velocity and
$\mu$ the mean molecular weight. In this case, Eq.\,(\ref{eq:settle})
becomes
\begin{equation}
  \frac{\partial(\rho L_j)}{\partial t}\bigg|_{\rm settle}\!\!\!
  ~=~ \frac{\partial}{\partial z}\Bigg(\bigg(\frac{3}{4\pi}\bigg)^{1/3}
    g\,\frac{\rho_{\rm m}}{\rho\,\vth}\;
    \sum_{i=1}^2 n_i\,V_i^{\;\frac{j}{3}+\frac{1}{3}}\Bigg)
  ~=~ \xi_{\rm lKn}\,\frac{\partial}{\partial z}
    \bigg(\frac{\rho_{\rm m}}{c_T}\,L_{j+1}\bigg) \ ,
\end{equation}
where $\xi_{\rm
  lKn}\!=\!g\,\big(\frac{3}{4\pi}\big)^{1/3}\frac{\sqrt{\pi}}{2}$ and
$c_T\!=\!\big(\frac{2\,kT}{\mu}\big)^{1/2}$. This is exactly the
result for settling obtained by \cite{Woitke2020}, see their
Eq.\,(25), which means that using the double-$\delta$
representation (Eq.\,\ref{eq:double-d}) leads to the same result as
working out the integral over size space.

\subsection{Small Knudsen numbers (Stokes regime)}
\noindent
For small Knudsen numbers, the drift motion of the particles becomes
viscous, see Eqs.\,(14), (15) and (16) in \cite{Woitke2003}.  For
small Reynolds numbers $Re = 2a\,\rho\,\vdreq/\nu_{\rm dyn} \la 1$ the
flow of the gas around the particle is laminar and the formula for
Stokes friction applies
\begin{equation}
  \vdreq = g\,\frac{2\,\rho_{\rm m}\,a^2}{9\,\nu_{\rm dyn}} \ ,
\end{equation}
where $\nu_{\rm dyn}$ is the dynamic gas viscosity $\rm[g/cm/s]$, which
we calculate for a (97\,\% \ce{CO2}, 3\,\% \ce{N2})-mixture according
to Eq.\,(8) in \cite{Woitke2003} with collisional molecular radii from
Table~1 in \cite{Woitke2022}, resulting in
\begin{equation}
  \nu_{\rm dyn} = 1.48\times 10^{-4}{\rm\frac{g}{cm\,s}}
                \,\left(\frac{T}{\rm 300\,K}\right)^{0.5} \ .
  \label{eq:viscosity}
\end{equation}
In the Stokes regime we find
\begin{equation}
  \frac{\partial(\rho L_j)}{\partial t}\bigg|_{\rm settle}\!\!\!
  ~=~ \frac{\partial}{\partial z}\Bigg(\bigg(\frac{3}{4\pi}\bigg)^{2/3}
    g\,\frac{2\,\rho_{\rm m}}{9\,\nu_{\rm dyn}}\;
    \sum_{i=1}^2 n_i\,V_i^{\;\frac{j}{3}+\frac{2}{3}}\Bigg)\\
  ~=~ \xi_{\rm sKn}\,\frac{\partial}{\partial z}
    \bigg(\frac{\rho_{\rm m}}{\nu_{\rm dyn}}\,\rho L_{j+2}\bigg) \ ,
\end{equation}
where $\xi_{\rm sKn}=g\,\frac{2}{9}\Big(\frac{3}{4\pi}\Big)^{2/3}$.
This is exactly the result for settling in \cite{Woitke2003}, see
their Eqs.\,(74) and (76).

\begin{figure*}
  \centering
  \begin{tabular}{ccc}
  \hspace*{-6mm}
  \includegraphics[page=1,height=77mm,trim=0 22 0 15,clip]
                  {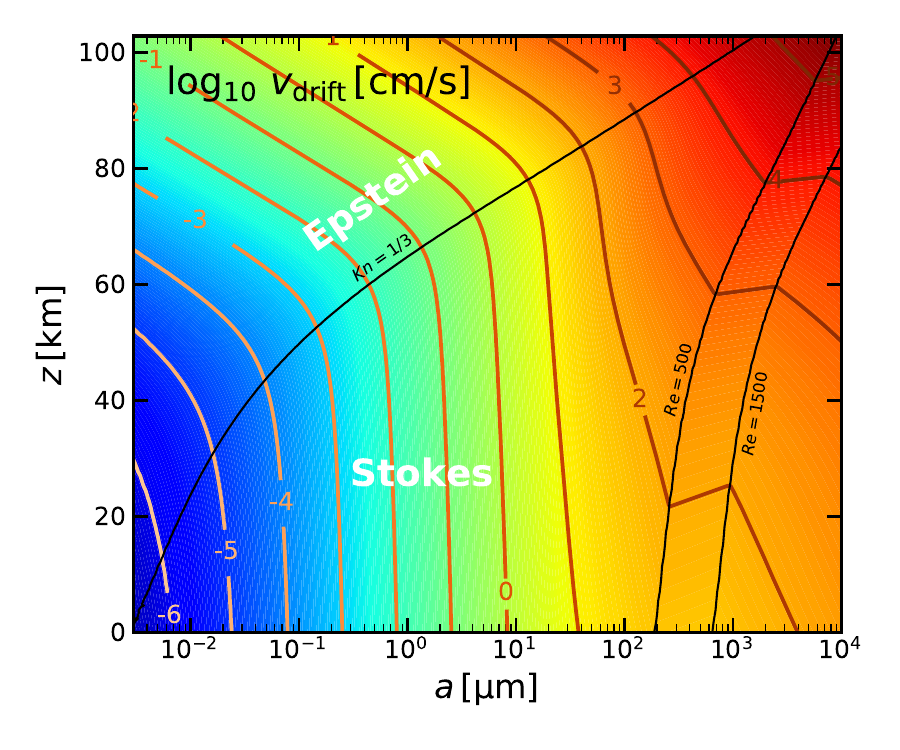} &
  \hspace*{-9mm}
  \includegraphics[page=2,height=77mm,trim=61 22 0 15,clip]
                  {vdrift.pdf}             
  \end{tabular}
  \caption{{\bf Left:} Equilibrium drift velocity $\vdreq$ (or final
    fall speed) of particles as function of particle radius $a$ and
    height $z$ in the Venus atmosphere, where we have assumed a
    material density of $\rho_{\rm m}\!=\!1.83\rm\,g/cm^3$, the mass
    density of sulphuric acid \ce{H2SO4}. {\bf Right:} The growth
    timescale by condensation of sulphuric acid on a particle's
    surface $\tau_{\rm gr} = V\,/\,(\partial V/\partial t)$ in
    the Venus atmosphere. Here we have assumed a large supersaturation
    $S\!\to\!\infty$ and \ce{H2O} to be the key molecule that limits
    the growth with a constant concentration of 30\,ppm. The blue
    dashed line shows where the growth timescale equals the sinking
    timescale $\tau_{\rm sink}\!=\!H_p/\vdreq$. Liquid
    \ce{H2SO4} is expected to evaporate below $45-50\,$km
    ($\sim\!350\,$K), see \cite{Titov2018}.}
  \label{fig:settle_growth}
\end{figure*}

\subsection{General Knudsen numbers}
\label{AppA3}
\noindent
In the general case, which is now implemented in {\sc DiffuDrift v2},
we use Eqs.\,(12), (14), (15) and (18) in \cite{Woitke2003} to
calculate the general equilibrium drift velocities $\vdreq_{,i}$ of both
representative particles, and then use Eq.\,(\ref{eq:settle}) to
calculate the effect of the settling on the cloud particle moments.
This method only becomes approximate when the two representative
particles are not in the same hydrodynamical domain, otherwise this
method is exact.  Figure \ref{fig:settle_growth} shows
$\vdreq(a,z)$ in the Venus atmosphere, where we have taken $\rho(z)$
and $T(z)$ from our underlying atmospheric structure assumed for
Venus, and assume a gas consisting of 97\% \ce{CO2} and 3\% \ce{N2},
see Fig.~\ref{fig:ggchem}. These results are in excellent agreement
with Fig.~3 in \cite{Seager2021}. The figure shows the
proportionality $\vdreq\!\propto\!a/\rho$ in the Epstein and
$\vdreq\!\propto\!a^2$ in the Stokes regime.  In the Stokes regime,
the equilibrium drift velocity is height-independent, and $10\times$
larger particles fall $100\times$ faster.  The transition between the
Stokes and the Epstein regime takes place at about
$Kn = 2a/\ell \approx 1/3$, where $\ell$ is the mean free path of
the molecules in the atmosphere.  For Reynolds numbers between 500 and
1500, the viscous flow of the gas around the particles changes from
laminar to turbulent, which temporarily reduces the frictional force
and the fall speed.  This effect plays a role for particles
$\ga\!100\,\mu$m, similar to rain droplets in the Earth atmosphere.

\section{Growth and evaporation}
\label{AppB}
\noindent
During a small time step $\Delta t$, the volume of a particle $i$
changes by a small increment $\Delta V_i\!\ll\!V_i$ via deposition and
sublimation of molecules on/from the surface of the particle, which we
call growth and evaporation. Using the double-$\delta$ representation
of the cloud particle size distribution function $f(V)$
(Eq.\,\ref{eq:double-d}), the corresponding change of the cloud
particle moments is
\begin{eqnarray}
  \Delta(\rho L_j)
  &~=~&
  \sum_{i=1}^2 n_i\,\Big(\big(V_i+\Delta V_i\big)^{j/3}-V_i^{\,j/3}\Big) 
  ~\approx~ \frac{j}{3} \sum_{i=1}^2 n_i V_i^{\,j/3-1} \Delta V_i
  \nonumber\\
  \Rightarrow&\hspace*{-4mm}&
  \frac{\partial(\rho L_j)}{\partial t}\Big|_{\rm growth}
  ~=~ \frac{j}{3} \sum_{i=1}^2 n_i V_i^{\;j/3-1}\,
      \frac{\partial V_i}{\partial t} \ .
  \label{eq:growth}
\end{eqnarray}
In order to use Eq.\,(\ref{eq:growth}), we need to know the growth
rates $\frac{\partial V_i}{\partial t}$ of both representative
particles $i\!=\!1$ and $i\!=\!2$. In \cite{Woitke2003} we have
derived expressions for $\partial V/\partial t$ in the Epstein
and the Stokes regimes.

%two relevant hydrodynamical regimes: (a) large Knudsen numbers
%$Kn=\ell/(2a)$ and subsonic drift velocities, also called the Epstein
%regime, and (b) small $Kn$ and small particle Reynolds numbers
%$Re=2a\,\vdreq/\nu_{\rm dyn}$, called the Stokes regime. We also
%discussed how to smoothly go from one to the other limiting case based
%on $Kn$. $\ell$ is the mean free path length, $a$ the particle's
%radius, and $\nu_{\rm kin}$ the kinematic gas viscosity $\rm[cm^2/s]$.

%In the following, we show that using the representative particle sizes
%with Eq.\,(\ref{eq:growth}) leads to the same result results as
%solving the integrals in both $Kn$ number limiting cases, which
%underlines that Eq.\,(\ref{eq:double-d}) is indeed a valid approach.

\subsection{Large Knudsen numbers (Epstein regime):}
\noindent
For large Knudsen numbers, the impinging molecules reach the 
particle from the distant gas in a single flight according to a
Maxwellian velocity distribution. 
\begin{equation}
  \frac{\partial V}{\partial t} =
  4\pi a^2 \sum_r n_r^{\rm key} v^{\rm rel}_r \alpha_r V_r \, \gamma_r
  \label{eq:growth_lKn}
\end{equation}
with the saturation factor
\begin{equation}
  \gamma_r = \left(1-\frac{1}{\Sr}\right)\;\times\;
  \left\{\begin{array}{cl}
  1 &               \mbox{\!\!if $\Sr\geq 1$}\\
  \!\!b_{\rm mix}^s & \mbox{\!\!if $\Sr< 1$} \ ,
  \end{array}\right. 
\end{equation}
which is positive when the condensate is supersaturated (causing
deposition), but becomes negative when the condensate is
undersaturated (causing sublimation). $r$ is an index for the included
surface reactions. $n_r^{\rm key}$ is the density of the
growth-limiting key reactant of the surface reaction, $v_r^{\rm
  rel}\!=\!\big(\frac{kT}{2\pi\,m_r}\big)^{1/2}$ its thermal relative
velocity, $m_r$ is the mass of the key species, and $\alpha_r$ is a
sticking probability.  $V_r$ is the condensed volume
created per sticking key molecule. $S_{\!r}$ is the reaction supersaturation
ratio, which in the most simple case equals the supersaturation ratio
$S$ of the material affected by reaction $r$.  $b_{\rm
  mix}^s\!=\!L_3^s/L_3$ is the volume mixing ratio of material
$s$ in the particle, which is only relevant when $S_{\!r}\!<\!1$,
assuming that the sublimation only occurs from patches on the surface
made of material $s$. Equation (\ref{eq:growth_lKn}) is valid in the
Epstein regime, for which the moment method was originally developed
by \cite{Gail1988} and later adopted for brown dwarfs and exoplanet
atmospheres by \cite{Helling2006,Helling2008}.  In this case,
Eq.\,(\ref{eq:growth}) results in
\begin{equation}
  \frac{\partial(\rho L_j)}{\partial t}\bigg|_{\rm growth}\!\!\!
  ~=~ \frac{j}{3}\;4\pi\,\bigg(\frac{3}{4\pi}\bigg)^{2/3}
  \sum_r n_r v^{\rm th}_r \alpha_r V_r\,\gamma_r
  \sum_{i=1}^2 n_i\,V_i^{\frac{j}{3}-1+\frac{2}{3}} 
  ~=~ \frac{j}{3}\,\chi^{\rm net}_{\rm lKn}\;\rho L_{j-1} \ ,
\end{equation}
where $\chi^{\rm net}_{\rm lKn}\!=\!\sqrt[3]{36\pi} \sum_r n_r v^{\rm
  th}_r \alpha_r V_r\,\gamma_r$, which is exactly the result for the
growth as formulated by Eqs.\,(11) and (12) in \cite{Woitke2020}.

\subsection{Small Knudsen numbers:}
\noindent
In the case of small Knudsen numbers, the molecules approaching a
particle undergo multiple collisions before they reach the surface,
and the transport of the molecules to the particle's surface becomes
diffusion-limited, see Eq.~(32) in \cite{Woitke2003}.
\begin{equation}
  \frac{\partial V}{\partial t}
  = 4\pi a \sum_r n_r^{\rm key} D_r V_r \,\gamma_r \ .
\end{equation}
Here, $D_r$ is the bi-molecular diffusion constant $\rm[cm^2/s]$ of the key
molecule of mass $m_r$ in the ambient gas with mean molecular weight $\mu$,
for which use
\begin{equation}
  D_r = 1.52\times 10^{18} \times
  \bigg(\frac{\rm amu}{m_r} + \frac{\rm amu}{\mu}\bigg) \frac{\sqrt{T}}{n}
\end{equation}
from \cite{Banks1973}, where $n\!=\!p/kT$ is the total gas particle
density. We note that since $D_r\!\propto\!1/n$, the growth rate
is independent of the gas density for small Knudsen numbers and hence
roughly independent of height for small particles in the lower Venus
atmosphere. The result for Eq.\,(\ref{eq:growth}) in this case is
\begin{equation}
  \frac{\partial(\rho L_j)}{\partial t}\bigg|_{\rm growth}\!\!\!
  ~=~ \frac{j}{3}\;4\pi\,\bigg(\frac{3}{4\pi}\bigg)^{1/3}
  \sum_r n_r^{\rm key} D_r V_r\,\gamma_r\,
  \sum_{i=1}^2 n_i\,V_i^{\frac{j}{3}-1+\frac{1}{3}} 
  ~=~ \frac{j}{3}\,\chi^{\rm net}_{\rm sKn}\;\rho L_{j-2} \ ,
\end{equation}
where $\chi^{\rm net}_{\rm sKn}\!=\!\sqrt[3]{48\pi^2}
\sum_r n_r^{\rm key} D_r V_r\,\gamma_r$, which is exactly the result
given by Eq.\,(75) in \cite{Woitke2003} when generalised to a mixture
of condensates.

\subsubsection{Rapidly falling particles}
\noindent
For very large particles, the equilibrium drift may become so fast
that the particles sweep up more molecules as they fall than they can gain
by diffusion.  We account for this effect by an additional term in the
volume growth rate for small Knudsen numbers
\begin{equation}
  \frac{\partial V}{\partial t}
  = \sum_r n_r^{\rm key}\,\Big(4\pi\,a\,D_r + \pi a^2\,\vdreq\Big)
    \,V_r\,\gamma_r \ .
  \label{eq:growth_sKn}
\end{equation}

\subsection{General Knudsen numbers}
\noindent
For general Knudsen numbers, as now implemented in {\sc DiffuDrift
v2}, we use Eqs.\,(\ref{eq:growth_lKn}) and (\ref{eq:growth_sKn}) to
calculate the growth rates $\partial V_i/\partial t$ for the two
representative particles in the large and in the small Knudsen number
limiting cases, and then use Eq.\,(36) in \cite{Woitke2003} to merge
them.  Next, we use Eq.\,(\ref{eq:growth}) to calculate the effect of
these general growth rates on the cloud particle moments. As for the
settling, this procedure only becomes approximate when both
representative particles are not in the same hydrodynamical domain.

Figure \ref{fig:settle_growth} shows the growth timescale $\tau_{\rm
  gr} = V\,/\,(\partial V/\partial t)$ for sulphuric acid droplets
in the Venus atmosphere when the sulphuric acid is highly
supersaturated and the growth is limited by the concentration of water
vapour (assuming a concentration of 30\,ppm). We obtain a pattern as
function of particle radius $a$ and height $z$ in the atmosphere that
is very similar to the drift velocity.

The blue dashed line shows where the growth timescale equals the
sinking timescale $\tau_{\rm sink}\!=\!H_p/\vdreq$. To the left of
this line, the particles have plenty of time to grow until saturation
is reached in the gas phase, which slows down further growth, before they
fall down.  Particles to the right of this line should not exist,
because they fall quicker than they can grow, however coagulation
might help them to cross this line temporarily.

\section{Coagulation and Charges}
\label{AppC}
\noindent
Our implementation of coagulation uses a discretised form of the
Smoluchowski equation \citep{Smoluchowski1916}. When particles with
volumes $V_1$ and $V_2$ collide and stick together, there are three
additional sizes of particles after a small time interval, namely
$V_3\!=\!V_1+V_1$, $V_4\!=\!V_1+V_2$, and $V_5\!=\!V_2+V_2$. Let
\begin{equation}
  \Coll_{ij} = \alpha_{ij}\;\pi (a_i+a_j)^2\,\Delta v_{ij}
  \;f^{\rm C}_{i,j}\;n_i\,n_j\,\Delta t
\end{equation}
denote the numbers of collisions that take place between particles of
sizes $i$ and $j$ during $\Delta t$ per cm$^3$, where $\alpha_{ij}$ is
the sticking probability, $\Delta v_{ij}$ the mean relative
velocity between the two particles, $a_i$ and $a_j$ their radii
($a\!=\big(3V/(4\pi)\big)^{1/3}$), $f^{\rm C}_{i,j}$ is the
Coulomb factor for electrostatic repulsion, and $n_i$ and $n_j$ are the
particles' number densities. Then
\begin{eqnarray}
  n_1' &\,=\,& n_1 - 2\,\Coll_{11} - \Coll_{12} \\
  n_2' &=& n_2 - 2\,\Coll_{22} - \Coll_{12} \\
  n_3' &=& \Coll_{11} \\
  n_4' &=& \Coll_{12} \\
  n_5' &=& \Coll_{22}
\end{eqnarray}
are the particle number densities after $\Delta t$.  This scheme can
easily be extended to $i\!=\!1,...\,,N$ particle sizes, in which case
$i\!=\!1,...\,,N'$ sizes of particles occur after a small $\Delta t$,
where $N'\!=\!N+N(N+1)/2$.  The change of the cloud particle moments
by coagulation is then calculated as
\begin{align}
  & \frac{\partial(\rho L_j)}{\partial t}\Bigg|_{\rm coag} \!\!=~
    \frac{1}{\Delta t}\,\Big(L_j(t+\Delta t)-L_j(t)\Big) \\[-1mm]
  & \mbox{where}~~~
  \rho L_j(t) = \sum_{i=1}^N n_i  V_i^{\,j/3}
  ~~\mbox{and}~~
  \rho L_j(t+\Delta t) = \sum_{i=1}^{N'} n_i' V_i^{\,j/3} 
  \nonumber
\end{align}
Concerning the relative velocity $\Delta v_{ij}$ we take into account
three processes: Brownian motion, difference in settling velocities,
and turbulence
\begin{eqnarray}
  \Delta v_{ij} &\,=\,& 
  \Big((v_{ij}^{\rm\,Brown})^2 + (v_{ij}^{\rm\,sett})^2
                           + (v_{ij}^{\rm\,turb})^2\Big)^{1/2} \\
  v_{ij}^{\rm\,Brown} &=&
  \left(\frac{8\,kT}{\pi\,m_{\rm red}}\right)^{1/2} \\
  v_{ij}^{\rm\,sett} &=&
  \Big|\,\vdreq_{,i}-\vdreq_{,j}\Big| \\
  v_{ij}^{\rm\,turb} &=& v_{ij}^{\rm\,turb}\,(V_g,Re,\St_i,\St_j) \ ,
\end{eqnarray}
where $m_{\rm red}\!=\!m_i\,m_j/(m_i+m_j)$ is the reduced mass of the
two colliding particles.  If the colliding particles carry charges of
equal sign, $q_i$ and $q_j$, the Coulomb factor for electrostatic
repulsion is
\begin{equation}
  f^{\rm C}_{i,j} \,=\, \exp\big(-E_{\rm C}/E_{\rm kin}\big)
  \label{eq:frep}
\end{equation}  
where $E_{\rm C} = q_i\,q_j\,{\rm e}^2/(a_i+a_j)$ is the electrostatic
repulsion energy at contact distance, e is the CGS unit of charge, and
$E_{\rm kin} = \frac{1}{2}\,m_{\rm red}\,\Delta v_{ij}^2$ is the
collisional kinetic energy.  Equation~(\ref{eq:frep}) can be derived
from elemental assumptions about the collisions between charged
particles when they have a Maxwellian velocity distribution (Brownian
motion); it states that coagulation becomes very inefficient when
$E_{\rm C}$ exceeds a few $E_{\rm kin}$.

\cite{Balduin2023} developed a detailed astrochemical model for dust
charging in protoplanetary discs, taking into account cosmic ray
ionisation, electron attachment, charge exchange with molecular ions,
and photoeffect.  They found that grains of different sizes in the
disc midplane, which is entirely shielded from UV-photons, all obtain
about the same negative charge relative to size $-q/a$, of the order
of a few tens to a few hundreds of negative charges per micrometer of
grain radius $a$.  Since the grains in the model are mainly charged by
collisions with electrons and molecular cations, this result is easy
to understand. The electron attachments are faster than the charge
exchange reactions with molecular cations, so the particles start to
charge negatively, until their electric potential $E_{\rm C}={\rm
  e}^2(q/a)$ becomes larger than a few $kT$, and the electrons cannot
reach the grain surfaces anymore. This is where the charging process
stops, leading to an estimation of the particle charge as
\begin{equation}
  \frac{q}{a} \,\approx\,-\frac{5\,kT}{\rm e^2}
              \,\approx\,-\frac{90}{\mu\rm m}
              \left(\frac{T}{300\rm\,K}\right) \ ,
  \label{eq:charge}
\end{equation}
which is independent of the gas density and the cosmic ray ionisation
rate.  The grain charging causes the density of the free electrons to
decrease by orders of magnitude. The density of negative charges on
the grains is eventually balanced by the density of molecular cations
in the gas, such as \ce{NH4+}, which have so large proton affinities
that they cannot recombine when they collide with a negatively charged
grain. The cosmic ionisation rate of the gas is proportional to both
the electron density and the molecular ion density, but it is the
ratio of the two that sets the dust charge. We have applied this model
to the problem of aerosol particle charges at the bottom of the Venus
atmosphere and find values compatible with Eq.\,(\ref{eq:charge}).

If tribo-electric charging is important, as expected in highly
turbulent environments, particle charges have been observed to be even
larger.  Laboratory simulations of the electrification of wind-driven
Martian dust particles \citep{Merrison2012} suggest a charge of $10^4$
to $10^6$ elementary charges per grain. Similar charges per grain
($10^5$ to $10^6$) have been measured in experiments on volcano ash
particles in Icelandic volcanic plumes
\citep{Aplin2014,MendezHarper2021}, including fragmentation charging.
\cite{MendezHarper2024} have found similar levels of
tribo-electrification when grinding coffee grains.

Our {\sc DiffuDrift} models show that if Eq.\,(\ref{eq:charge})
was true, the aerosol particles in the lower Venus atmosphere
could not collide much, and hence the models with coagulation
very much resemble the models without coagulation. However, since
the charge of the aerosol particles in the lower Venus atmosphere
is not known exactly, we treat $q/a$ at $T=300\,$K as a free
parameter, see Eq.\,(\ref{eq:qa300}).

The turbulent relative velocity $v_{ij}^{\rm\,turb}$ is calculated
according to the vertical gas mixing velocity $V_g$ in the three cases
for small, medium and large particles according to Eqs.\,(26), (28)
and (29) in \cite{Ormel2007}, depending on the relations between the
stopping times $\tau_i^{\rm stop}$, $\tau_j^{\rm stop}$ of the
colliding particles and the smallest eddy turnover timescale
$\tau_\eta$.  The closed expressions in \cite{Ormel2007} are then
formulated in terms of the gas Reynolds number $Re$ and the particles'
Stokes numbers $\St_i$ and $\St_j$.
\begin{align}
  \mu  & && \mbox{mean molecular weight $\rm[g]$} \\
  H_p  & ~=~\frac{kT}{\mu\,g}
         && \mbox{atmospheric scale height $\rm[cm]$} \\
  V_g  & ~=~\frac{D}{H_p} \label{eq:Vg}
         && \mbox{vertical gas mixing velocity $\rm[cm/s]$} \\
  \nu_{\rm dyn} & \mbox{~~:~~Eq.\,(\ref{eq:viscosity})}
         && \mbox{dynamic gas viscosity $\rm[g/cm/s]$} \\
  v_{\rm th} & ~=~\left(\frac{8\,kT}{\pi\,\mu}\right)^{1/2}\!\!\!
        && \mbox{thermal velocity $\rm[cm/s]$} \\
  \ell & ~=~\frac{3\,\nu_{\rm dyn}}{\rho\,v_{\rm th}}
        && \mbox{mean free path $\rm[cm]$} \\
  Re   & ~=~\frac{V_g\,\rho\,H_p}{\nu_{\rm dyn}}
         && \mbox{gas Reynolds number}\\
  \tau_{\rm edd} & ~=~\frac{H_p}{V_g}
         && \mbox{largest eddy turnover timescale $\rm[s]$}\\
  \tau_\eta & ~=~\tau_{\rm edd}\,Re^{-1/2}\!\!\!\!
         && \mbox{smallest eddy turnover timescale $\rm[s]$}
\end{align}
\begin{align}
  \tau_i^{\rm stop} & ~=~\frac{\vdreq(a_i)}{g}
         && \mbox{stopping timescale $\rm[s]$} \\
  \St_i & ~=~\frac{\tau_i^{\rm stop}}{\tau_{\rm edd}}
         && \mbox{Stokes number}
\end{align}
We note that the turbulent relative velocity is limited by
$v_{ij}^{\rm\,turb}\la V_g$, and the mean vertical gas mixing
velocity $V_g$ is assumed to be related to the local eddy diffusion
coefficient $D$ via the general relation $D=V_g\,H_p$
(Eq.\,\ref{eq:Vg}).  Since $D$ is assumed to be quite small in the
Venus atmosphere (Eq.\,\ref{eq:eddy_diff}), and hence $V_g$ is less than
about 0.1\,cm/s, turbulence is found to be rather
unimportant. Instead, the relative collisional velocities $\Delta
v_{ij}$ are mostly driven by the Brownian motion, unless we have
millimetre-sized particles, in which case the difference in settling
velocities becomes more important.

The cloud particles are expected to remain quite small, less than
about 1\,mm (see $\tau_{gr}\!=\!\tau_{\rm sink}$ line in
Fig.\,\ref{fig:settle_growth}). Since the collisional difference
velocities $\Delta v_{i,j}$ are found to be $<10\,$cm/s throughout
the lower Venus atmosphere in our models, the collisions are expected
to be far from the bouncing and fragmentation barriers. We therefore
assume perfect sticking $\alpha_{i,j}\!=\!1$.

\end{appendix}

\end{document}